\pdfoutput=1
\documentclass{article}
\usepackage[a4paper,left=2.2cm,right=2.2cm,bottom=1.5cm,top=1.5cm,includefoot,includehead]{geometry}

\usepackage{mathtools}
\DeclarePairedDelimiter{\ceil}{\lceil}{\rceil}

\usepackage{siunitx}

\usepackage {tikz}

\usepackage{times}
\usepackage{graphicx} 
\usepackage{subcaption} 

\usepackage{natbib}

\usepackage{algorithm}
\usepackage{algorithmic}
\usepackage{xr}

\usepackage{hyperref}

\usepackage{amsmath}
\usepackage{amsthm}
\usepackage{amssymb}
\usepackage{mynotation} 
\usepackage{authblk}

\usepackage{xcolor,soul}
\newcommand{\modh}[1]{{#1}}

\usetikzlibrary{patterns}
\newlength{\hatchspread}
\newlength{\hatchthickness}
\newlength{\hatchshift}
\newcommand{\hatchcolor}{}
\tikzset{hatchspread/.code={\setlength{\hatchspread}{#1}},
	hatchthickness/.code={\setlength{\hatchthickness}{#1}},
	hatchshift/.code={\setlength{\hatchshift}{#1}},
	hatchcolor/.code={\renewcommand{\hatchcolor}{#1}}}
\tikzset{hatchspread=8pt,
	hatchthickness=0.4pt,
	hatchshift=0pt,
	hatchcolor=black}
\pgfdeclarepatternformonly[\hatchspread,\hatchthickness,\hatchshift,\hatchcolor]
{custom north west lines}
{\pgfqpoint{\dimexpr-2\hatchthickness}{\dimexpr-2\hatchthickness}}
{\pgfqpoint{\dimexpr\hatchspread+2\hatchthickness}{\dimexpr\hatchspread+2\hatchthickness}}
{\pgfqpoint{\dimexpr\hatchspread}{\dimexpr\hatchspread}}
{
	\pgfsetlinewidth{\hatchthickness}
	\pgfpathmoveto{\pgfqpoint{0pt}{\dimexpr \hatchspread+\hatchshift}}
	\pgfpathlineto{\pgfqpoint{\dimexpr\hatchspread+0.15pt+\hatchshift}{-0.15pt}}
	\ifdim \hatchshift > 0pt
	\pgfpathmoveto{\pgfqpoint{0pt}{\hatchshift}}
	\pgfpathlineto{\pgfqpoint{\dimexpr0.15pt+\hatchshift}{-0.15pt}}
	\fi
	\pgfsetstrokecolor{\hatchcolor}
	\pgfusepath{stroke}
}

\pgfdeclarepatternformonly[\hatchspread,\hatchthickness,\hatchshift,\hatchcolor]
{custom north east lines}
{\pgfqpoint{\dimexpr-2\hatchthickness}{\dimexpr-2\hatchthickness}}
{\pgfqpoint{\dimexpr\hatchspread+2\hatchthickness}{\dimexpr\hatchspread+2\hatchthickness}}
{\pgfqpoint{\dimexpr\hatchspread}{\dimexpr\hatchspread}}
{
	\pgfsetlinewidth{\hatchthickness}
	\pgfpathmoveto{\pgfqpoint{\dimexpr\hatchshift-0.15pt}{-0.15pt}}
	\pgfpathlineto{\pgfqpoint{\dimexpr\hatchspread+0.15pt}{\dimexpr\hatchspread-\hatchshift+0.15pt}}
	\ifdim \hatchshift > 0pt
	\pgfpathmoveto{\pgfqpoint{-0.15pt}{\dimexpr\hatchspread-\hatchshift-0.15pt}}
	\pgfpathlineto{\pgfqpoint{\dimexpr\hatchshift+0.15pt}{\dimexpr\hatchspread+0.15pt}}
	\fi
	\pgfsetstrokecolor{\hatchcolor}
	\pgfusepath{stroke}
}

\pgfdeclarepatternformonly[\hatchspread,\hatchthickness,\hatchshift,\hatchcolor]
{custom vertical lines}
{\pgfqpoint{\dimexpr-2\hatchthickness}{\dimexpr-2\hatchthickness}}
{\pgfqpoint{\dimexpr\hatchspread+2\hatchthickness}{\dimexpr\hatchspread+2\hatchthickness}}
{\pgfqpoint{\dimexpr\hatchspread}{\dimexpr\hatchspread}}
{
	\pgfsetlinewidth{\hatchthickness}
	\pgfpathmoveto{\pgfqpoint{\dimexpr\hatchshift-0.15pt}{-0.15pt}}
	\pgfpathlineto{\pgfqpoint{\dimexpr-0.15pt}{\dimexpr\hatchspread-\hatchshift-0.15pt}}
	\ifdim \hatchshift > 0pt
	\pgfpathmoveto{\pgfqpoint{-0.15pt}{\dimexpr\hatchspread-\hatchshift-0.15pt}}
	\pgfpathlineto{\pgfqpoint{\dimexpr\hatchshift+0.15pt}{\dimexpr\hatchspread+0.15pt}}
	\fi
	\pgfsetstrokecolor{\hatchcolor}
	\pgfusepath{stroke}
}

\title{Probabilistic modelling and reconstruction of strain}
\author[1]{Carl Jidling}
\author[2]{Johannes Hendriks}
\author[1]{Niklas Wahlstr\"{o}m}
\author[2]{Alexander Gregg}
\author[1]{Thomas B. Sch\"{o}n}
\author[2]{Christopher Wensrich}
\author[2]{Adrian Wills}
\affil[1]{Department of Information Technology, Uppsala University, Sweden}
\affil[2]{School of Engineering, University of Newcastle, Australia}

\date{}

\begin{document}
	\newcommand{\coverTitle}{Probabilistic modelling and reconstruction of strain}
\newcommand{\coverAuthors}{Carl Jidling, Johannes Hendriks, Niklas Wahlstr\"{o}m, Alexander Gregg, Thomas B. Sch\"{o}n, Christopher Wensrich and Adrian Wills}
\newcommand{\coverStatus}{Accepted for publication.}

\begin{titlepage}
	\begin{center}
		{\large \em Technical report}
		
		\vspace*{2.5cm}
		%
		{\Huge \bfseries \coverTitle  \\[0.4cm]}
		
		%
		{\Large \coverAuthors \\[2cm]}
		
		\renewcommand\labelitemi{\color{red}\large$\bullet$}
		\begin{itemize}
			\item {\Large \textbf{Please cite this version:}} \\[0.4cm]
			\large
			\coverAuthors. \coverTitle. \textit{Nuclear instruments and methods in physics research section B}, 436:141-155, 2018.	
		\end{itemize}
		
		\vfill
		
		\begin{abstract}
		This paper deals with modelling and reconstruction of strain fields, relying upon data generated from neutron Bragg-edge measurements.
We propose a probabilistic approach in which the strain field is modelled as a Gaussian process, assigned a covariance structure customised by incorporation of the so-called equilibrium constraints.
The computational complexity is significantly reduced
by utilising an approximation scheme well suited for the problem. 
We illustrate the method on simulations and real data.
The results indicate a high potential and can hopefully inspire the concept of probabilistic modelling to be used within other tomographic applications as well. 		\end{abstract}
		
		
		\vfill
	\end{center}
\end{titlepage}

	\maketitle
	
	\begin{abstract}
		This paper deals with modelling and reconstruction of strain fields, relying upon data generated from neutron Bragg-edge measurements.
		We propose a probabilistic approach in which the strain field is modelled as a Gaussian process, assigned a covariance structure customised by incorporation of the so-called equilibrium constraints.
		The computational complexity is significantly reduced
		by utilising an approximation scheme well suited for the problem. 
		We illustrate the method on simulations and real data.
		The results indicate a high potential and can hopefully inspire the concept of probabilistic modelling to be used within other tomographic applications as well. 
	\end{abstract}
	
	

	\section{Introduction}\label{sec:intro}
	The goal of tomographic reconstruction is to build a map of an unknown quantity within an object using information gained from irradiation experiments.
	A well known example of this is X-ray imaging, where the unknown quantity might be, for instance, the \modh{bone density} inside a human body.
	
	Each measurement provides information about the amount of intensity that the ray has lost when passing through the material.
	Of course, a single measurement does not uniquely define the interior.
	However, processing a large number of measurements taken from many different angles allows for an accurate reconstruction of the internal structure.   
	
	While techniques such as X-ray imaging and MRI are concerned with scalar fields, we are in this work considering the reconstruction of the strain field -- a second order tensor -- within a deformed material.
	This is a significantly harder problem as it is a multidimensional quantity at each point.
	For simplicity, we are restricting the analysis to a planar problem, but the extension to three dimensions follows the same procedure.   
	
	The development of accurate strain measuring techniques is motivated by applications within several fields. 
One field with perhaps especially exciting application potential is \textit{additive manufacturing}, which involves printing of three-dimensional metal structures.
For instance, this is of interest for developers of fuel nozzles \citep{Tremsin2016} and turbine blades \citep{Watkins2013} within the aerospace industry.

Several techniques enabling high-precision measurement of residual strain have been proposed in previous work. 
These are characterised as destructive, semi-destructive or non-destructive, where examples from each category includes slitting \citep{Prime2001}, ring-coring \citep{Standard2001} and diffraction \citep{Fitzpatrick2003,Noyan2013}, respectively.

Bragg-edge analysis \citep{Santisteban2002,Santisteban_Eng} is an alternative transmission-based approach aiming at reconstructing the entire three-dimensional strain-field. 
This is an important difference to the established techniques outlined above, including the diffraction-based strain tomography such as synchrotron X-ray measurements \citep{Korsunsky2011,Korsunsky2006}. 
The term Bragg-edge refers to rapid changes in the relative transmission rate, which are determined from Bragg's law and hence directly related to the wavelength \citep{Santisteban_Eng}.

Methods relying on the Bragg-edge  idea have seen a significant progress during recent years, and provides an essential foundation for generation of high-resolution strain-images within polycrystalline materials \citep{Santisteban2002,Tremsin2012,Tremsin2011,Woracek2018}.  

A practical method based on the Brag-edge technique has been proposed by \citet{Wensrich2016a}. 
This relies upon the assumption that the strain field is compatible (see Section D), which allows the measurement equation to be expressed in terms of the boundary displacements. 
The reconstruction is obtained by solving a least squares problem and providing the result as a boundary condition to a finite element solver. 
An extension of this method is given by \citet{Wensrich2016b} with application to a real-world problem presented in \citep{Hendriks2017}.
	
	
	The solution technique presented in this work relies upon the concept of \textit{probabilistic modelling} \citep{Zoubin2015}.
	Probabilistic modelling refers to methods that are employing probability theory to encode uncertainties present in the problem and where the solution is obtained through statistical inference. 
	The idea is based on the assumption that uncertainties are always present, mainly due to a limited amount of data and the presence of measurement noise.
	A natural way of encoding these uncertainties in the model is therefore to assign a probabilistic measure to the unknown quantities themselves.  
	
	The contribution of this paper is a new way of modelling and reconstructing strain fields from data generated by neutron Bragg-edge measurements.
	We are using a tailored Gaussian process (GP)\citep{Rasmussen2006} to model the strain field, and by utilising the fact that GPs are closed under linear transformations, the reconstruction of the strain field is obtained through GP regression. 
	The model is customised by designing the associated covariance function with respect to the so-called equilibrium constraints, which guarantees a physical solution.   
	
	  

\section{Problem formulation}\label{sec:probform}
	Given a set of measurements generated from a neutron Bragg-edge experiment, the problem faced in this work is to reconstruct the strain at various positions within a \textit{sample}. The sample is an object, in which we want to reconstruct strain. The sample is considered to be two-dimensional. The strain in such a sample can be represented using a symmetric $2\times 2$-matrix $\mat{\epsilon}$ called the \textit{strain tensor}. Any point in the sample has an assigned strain tensor. This assignment is described by the \textit{strain field} $\mat{\epsilon}(\x)$, which is a function mapping any point in space $\x = [\cx, \,\, \cy ]^\Transp\in\mathbb{R}^2$ to a strain tensor $\mat{\epsilon}$. The strain field can be construed as
	\begin{equation}\label{eq:strain_tensor}
		\mat{\epsilon}(\x) = 
		\begin{bmatrix}
		\epsilon_{\cx\cx}(\x) & \epsilon_{\cx\cy}(\x) \\ \epsilon_{\cy\cx}(\x) & \epsilon_{\cy\cy}(\x)
		\end{bmatrix}\in\mathbb{R}^{2\times2},
	\end{equation}
	where $\epsilon_{\cx\cy}(\x)=\epsilon_{\cy\cx}(\x)$ since strain tensors are symmetric.
	
\begin{figure}
		\centering
		\includegraphics[width=0.8\columnwidth]{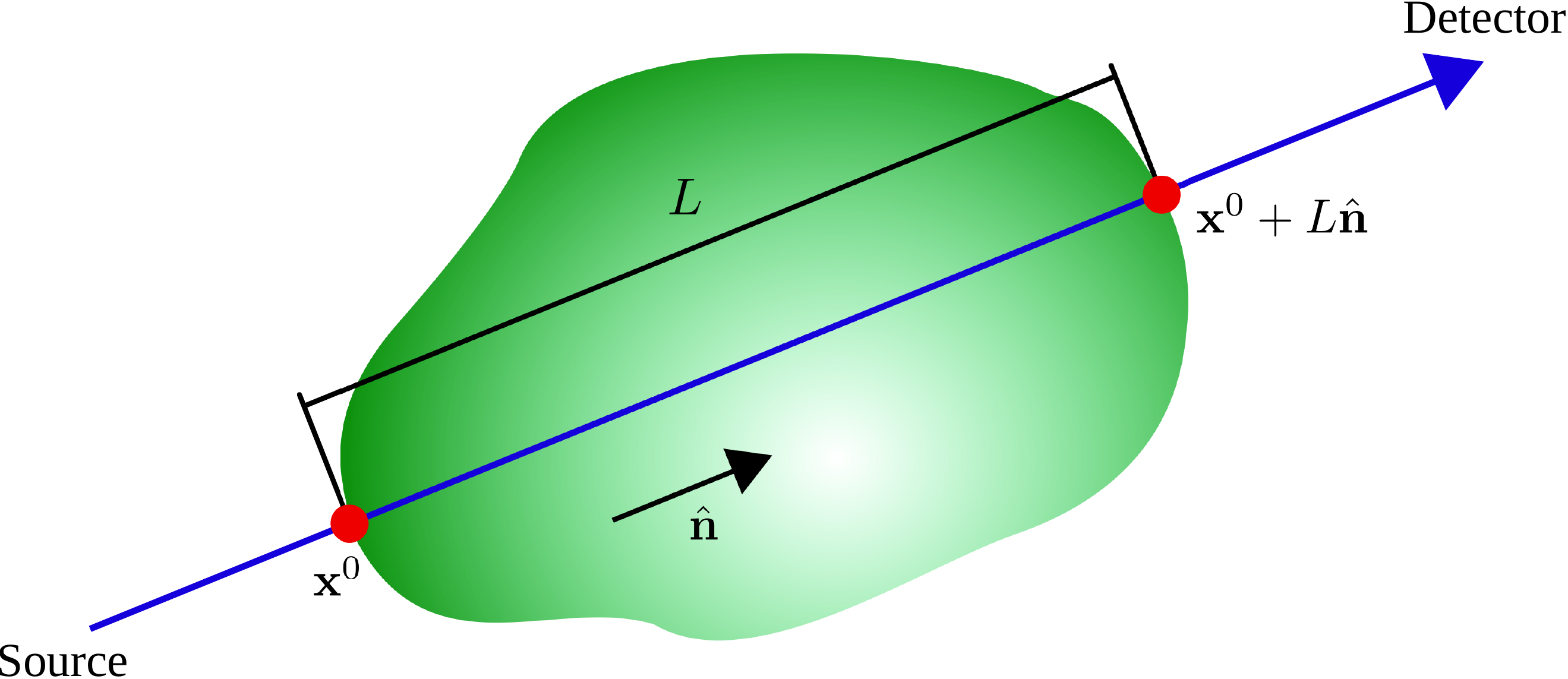}
		\caption{Experimental setup. The neutrons are transmitted from a source through the sample along the line starting at $\x^0$ and ending at $\x^0 + \hat \n L$ and finally measured by a detector.\label{fig:braggedge_setup} }
	\end{figure}

	The experiments rely upon high resolution time-of-flight neutron detectors. 
	Neutron beams are generated at a source, transmitted through the sample, and recorded at a detector located at the opposite side of the sample.
	Considering a single measurement, assume that the neutrons enters the sample at a point $\x^0$, propagates along the direction defined by the unit vector $\hat{n}$ and exits at $\x^0+L\hat{n}$, where $L$ is the illuminated distance in the sample.
	This is illustrated in Fig. \ref{fig:braggedge_setup}. 
	An ideal measurement obtained from the neutron Bragg-edge method can be expressed in terms of the \textit{Longitudinal Ray Transform} (LRT) 
	\begin{equation}\label{eq:main_int}
		\mathrm{I}(\vec{\eta}) =
		\frac{1}{L}\int_{0}^{L} \hat{n}^\Transp\mat{\epsilon}(\x^0+s\hat{n})\hat{n}ds,
	\end{equation}
	where $\vec{\eta}=\{\x^0,L,\hat{n}\}$ 
	specifies the argument of the LRT and where $s$ is a coordinate used to specify the position on the line between the entry and exit points.
	We can interpret \eqref{eq:main_int} as the average strain along the propagated path, so the LRT plays an important role in defining an adequate measurement model within this framework \citep{Lionheart2015}.
	See \ref{sec:BragEdgeMethod} for some more details on the Bragg-edge experiment. 
	
	In a real-world application, we will have access to a data set $\mathcal{D} = \{(\vec{\eta}_i,y_i)\}_{i=1}^N$ with
	\begin{equation} \label{eq:meas}
		y_i = \mathrm{I}(\vec{\eta}_i) + \varepsilon_i,
	\end{equation}
	where $\varepsilon_i$ denotes the measurement noise.
	Hence, the problem to be solved is this:
	Given the measurements $\{y_i\}_{i=1}^N$ in \eqref{eq:meas} and the relation \eqref{eq:main_int}, reconstruct the strain tensor \eqref{eq:strain_tensor} in any arbitrary \textit{test point} $\x_*$ along with an uncertainty measure of the accuracy for this reconstruction.


\section{Strain field reconstruction using Gaussian processes}\label{sec:braggedgeGP}
In this work, the problem is solved by making use of the framework known as \textit{probabilistic modelling},
with the somehow abstract idea of assigning a probability distribution to the strain field.
This does not mean that we should think of the strain field as a random object, but the distribution provides a way for us to encode the uncertainty of the reconstruction.
The presence of the uncertainty is mainly due to the fact that we have a finite number of measurements and that each of these measurements by construction are contaminated with noise. 

%

More specifically, to reconstruct the strain field based on LRT measurements \eqref{eq:main_int}, we need a probabilistic model describing (i) the strain field, and (ii) the relation between the strain field and the measurements. 
The model should be able to reason about the uncertainties present in the measurements and take known physical constraints into consideration. 
In this work we choose to model the strain field with a Gaussian process.

\subsection{Gaussian processes}
A GP is a stochastic process suitable for modelling spatially correlated measurements. GPs can be seen as a distribution over functions
\begin{align} \label{eq:GPprior}
\f(\x) 	&\sim \GP\big(\m(\x),\K(\x,\x')\big),
\end{align}
where the process is uniquely defined with its mean function $\m(\x) = \E[\f(\x)]$ and covariance function $\K(\x,\x') = \E\big[(\f(\x)-\m(\x))(\f(\x')-\m(\x'))^\Transp\big]$. The GP is a generalisation of the multivariate Gaussian probability distribution in the sense that the function values evaluated for a finite number of inputs $\x_1,\dots, \x_N$ are Gaussian distributed
\begin{subequations}
\label{eq:normalDist}
\begin{align} 
\begin{bmatrix} 
\f(\x_1)\\ 
\vdots \\ 
\f(\x_N)
\end{bmatrix} 	
&\sim \N(\vec{\mu},\K),
\quad
\text{where}
\quad
\vec{\mu}=
\begin{bmatrix} 
\m(\x_1)\\ 
\vdots \\ 
\m(\x_N)
\end{bmatrix},
\intertext{and}
\K&=
\begin{bmatrix}
\K(\x_1,\x_1) & \cdots & \K(\x_1,\x_N) \\
\vdots			&					& \vdots 			\\
\K(\x_N,\x_1) & \cdots & \K(\x_N,\x_N)
\end{bmatrix}.
\end{align}
\end{subequations}
In this work we will only consider zero-mean GPs, i.e., where $\m(\x) = \vec{0}$.
\modh{This is the natural choice in absence of more specific prior knowledge. 
It shall not be interpreted as if we believe that $\f(\x)=\vec{0}$, it simply represents the fact that we do not have any better initial guess.}

Since $\mat{\epsilon}(\x)$ is a symmetric $2\times2$ tensor, it consists of three unknown components, $\epsilon_{\cx\cx}(\x)$, $\epsilon_{\cx\cy}(\x)$ and $\epsilon_{\cy\cy}(\x)$. We therefore choose to model the strain tensor with a function $\f(\x): \mathbb{R}^2 \mapsto \mathbb{R}^3$, here called the \textit{strain function}, where
\begin{align}\label{eq:target_func}
\f(\x) = 
\begin{bmatrix} 
f_{\cx\cx}(\x)\\ 
f_{\cx\cy}(\x)\\ 
f_{\cy\cy}(\x)\\
\end{bmatrix}
=
\begin{bmatrix} 
\epsilon_{\cx\cx}(\x)\\ 
\epsilon_{\cx\cy}(\x)\\ 
\epsilon_{\cy\cy}(\x)\\
\end{bmatrix} \qquad \text{and} \qquad
\x = 
\begin{bmatrix}
\cx \\
\cy
\end{bmatrix}.
 \end{align}
 We then put a GP prior on $\f(\x)$ according to \eqref{eq:GPprior}. The model now consist of two parts:
 \begin{enumerate}
 	\item The GP prior \eqref{eq:GPprior} of the strain field, i.e., our choice of $\K(\x,\x')$. This GP prior is described in Section~\ref{sec:GPmodel}.
 	\item The joint distribution between the measurements $\y = [y_1, \, y_2, \, \dots, y_N]^\Transp$ and the strain function $\f_* = \f(\x_*)$ at a point $\x_*$ where we want to make a reconstruction. This distribution allows us to infer the measurements and it is described in Section~\ref{sec:GPmeas} and Section~\ref{sec:GPmeasExt}.
 \end{enumerate}

\subsection{The covariance function}
\label{sec:GPmodel}
The strain function cannot be any arbitrary function mapping $\mathbb{R}^2$ to $\mathbb{R}^3$. It needs to obey some physical laws. Therefore, we want to model the covariance function $\K(\x,\x')$ such that any sample we draw from the GP prior \eqref{eq:GPprior} is a valid strain function. More specifically we need to fulfil the equilibrium constraints, which for isotropic linearly elastic \modh{(i.e. in the absence of strong grain texture)} solid materials under the assumption of plane stress, read as 
		\begin{subequations}\label{eq:equi_constr_stress1}				
			\begin{align} 
			\frac{\partial f_{\cx\cx}(\x)}{\partial x} +
			(1-\nu)\frac{\partial f_{\cx\cy}(\x)}{\partial y}	+		
			\nu\frac{\partial f_{\cy\cy}(\x)}{\partial x} = 0, \\
			\nu\frac{\partial f_{\cx\cx}(\x)}{\partial y} +
			(1-\nu)\frac{\partial f_{\cx\cy}(\x)}{\partial x}	+		
			\frac{\partial f_{\cy\cy}(\x)}{\partial y} = 0,
			\end{align}
		\end{subequations}
		where $\nu$ denotes Poisson's ratio. For $\f(\x)$ to fulfill these constraints we can describe it as a transformation of another scalar function $\varphi(\x):\, \mathbb{R}^3 \mapsto \mathbb{R}$ via the transfomation according to
		\begin{equation}\label{eq:calL1}
		\f(\x)
		=
		\begin{bmatrix}
		\pdd{}{y}-\nu\pdd{}{x} \\[1mm]
		(1+\nu)\pdt{}{x}{y} \\[1mm]
		\pdd{}{x}-\nu\pdd{}{y}
		\end{bmatrix}\varphi(\x)
		=\mathcal{L}_\x\varphi.
		\end{equation}
		Here, $\mathcal{L}_\x$ is an \textit{operator} mapping scalar potential functions $\varphi(\x)$ to vector-valued strain functions $\f(\x)$, i.e., in this case we have $\mathcal{L}_\x: (\mathbb{R}^2 \mapsto \mathbb{R}) \mapsto (\mathbb{R}^2 \mapsto \mathbb{R}^3)$.
		One can easily verify that the strain function in \eqref{eq:calL1} fulfils the constraints in \eqref{eq:equi_constr_stress1}. 
		This transformation is derived from the underlying physics where $\varphi(\x)$ is a known physical potential called the \textit{Airy stress function}, for details, see \ref{sec:AiryStress}. 
		We can also derive the transformation $\f(\x)	
		=\mathcal{L}_\x\varphi$ from any set of linear constraints, for example those presented in \eqref{eq:equi_constr_stress1}, following the procedure described by \citet{Jidling2017}. 

		Instead of designing a covariance function for the strain function $\f(\x)$, we design a GP prior for the scalar potential function $\varphi(\x)$
		\begin{align} \label{eq:GPphi}
		\varphi(\x) 	&\sim \GP\big(0,k_\varphi(\x,\x')\big).
		\end{align}
		It can be easily verified in \eqref{eq:calL1} that the operator $\mathcal{L}_\x$ is linear 
		\[\mathcal{L}_\x[\gamma_1\varphi_1(\x)+\gamma_2\varphi_2(\x)] = \gamma_1\mathcal{L}_\x[\varphi_1(\x)]+\gamma_2\mathcal{L}_\x[\varphi_2(\x)].\]
		Since the GP $\varphi(\x)$ is mapped through this linear operator, it follows from \ref{sec:func_observs} that $\f(\x)$ is also a GP
		\begin{align} \label{eq:GPf}
		\f(\x) \sim
		\GP(\bs{0},
		\K(\x,\x')), 
		\intertext{where}
		\K(\x,\x') = 
		\mathcal{L}_\x k_\varphi(\x,\x')\mathcal{L}_{\x'}^\Transp.
		\end{align} 
		Any sample drawn from the GP prior \eqref{eq:GPf} will by this design obey the equilibrium constraints \eqref{eq:equi_constr_stress1} and hence represent a valid strain field.
		There are a variety of options for the scalar covariance function $k_\varphi$.
		The most common one is the so-called \textit{squared exponential} covariance function
		\begin{align}\label{eq:se_covfunc}
		k_\varphi(\vec{x,x'})=\sigma_f^2\exp\left[{-\frac{1}{2}(l_x^{-2}r_x^2+l_y^{-2}r_y^2)}\right],
		\end{align}
		where
		$r_x=x-x'$ and
		$r_y=y-y'$.  
		Here, $\sigma_f$ is a magnitude parameter while $l_x$ and $l_y$ determine the rate at which the covariance decays in direction $x$ and $y$, respectively. 
		These so-called \textit{hyperparameters} are learnt from data, more on this in Section \ref{sec:hypparams}.  
		
		Note that we can write 
		\[\K(\x,\x')=\mathcal{L}_\x k_\varphi(\x,\x')\mathcal{L}_{\x'}^\Transp=
		\mathcal{L}_\x\mathcal{L}_{\x'}^\Transp k_\varphi(\x,\x')
		=\bs{\Psi} k_\varphi(\x,\x'),\]
		where $\bs{\Psi}$ is a matrix of operators.
		Specifically,
		\begin{subequations}
			\begin{align}
			\bs{\Psi}_{11} & = \nu^2\frac{\partial^4}{\partial x^2\partial x'^2}
			-2\nu\frac{\partial^4}{\partial x^2\partial y'^2}
			+\frac{\partial^4}{\partial y^2\partial y'^2} ,
			\\
			\bs{\Psi}_{22} & =(\nu+1)^2\frac{\partial^4}{\partial x\partial y\partial x'\partial y'}   ,                                                                                           
			\\
			\bs{\Psi}_{33} & = \frac{\partial^4}{\partial x^2\partial x'^2}
			-2\nu\frac{\partial^4}{\partial x^2\partial y'^2}
			+\nu^2\frac{\partial^4}{\partial y^2\partial y'^2} ,
			\\
			\bs{\Psi}_{12} & =\bs{\Psi}_{21}=
			-\nu(\nu+1)\frac{\partial^4}{\partial x\partial y\partial x'^2}                                                                                                             
			+(\nu+1)\frac{\partial^4}{\partial x\partial y\partial y'^2},
			\\
			\bs{\Psi}_{13} & = \bs{\Psi}_{31} = 
			-\nu\frac{\partial^4}{\partial x^2\partial x'^2}
			+(\nu^2+1)\frac{\partial^4}{\partial x^2\partial y'^2}                                                     
			-\nu\frac{a}{b}\frac{\partial^4}{\partial y^2\partial y'^2},
			\\
			\bs{\Psi}_{23} & = \bs{\Psi}_{32} =
			-\nu(\nu+1)\frac{\partial^4}{\partial x\partial y\partial y'^2}
			+(\nu+1)\frac{\partial^4}{\partial x\partial y\partial x'^2}.
			\end{align}
		\end{subequations}
		For example, if we let $k_\varphi(\x,\x')$ be the squared exponential covariance function \eqref{eq:se_covfunc} we get
		 \begin{subequations}
		 	\begin{align}
		 	\frac{\partial^4}{\partial x^2\partial x'^2}k_\varphi&=   
		 	l_x^{-4}(l_x^{-4}r_x^4-6l_x^{-2}r_x^2+3)k_\varphi,
		 	\\
		 	\frac{\partial^4}{\partial x^2\partial y'^2}k_\varphi&=   
		 	\frac{\partial^4}{\partial x\partial y\partial x'\partial y'}k_\varphi
		 	l_x^{-2}l_y^{-2}(1-l_x^{-2}r_x^2)(1-l_y^{-2}r_y^2)k_\varphi,
		 	\\
		 	\frac{\partial^4}{\partial x\partial y\partial y'^2}k_\varphi&=   
		 	l_x^{-2}l_y^{-4}r_x r_y(l_x^{-2}r_y^2-3)k_\varphi,
		 	\end{align}
		 \end{subequations}
		with the remaining expressions obtained by exchanging $x\leftrightarrow y$. 
		
\subsection{The measurement model}
\label{sec:GPmeas}
In the previous subsection we presented a GP model for the strain function stipulating that the strain tensors at any two points will be jointly Gaussian distributed. 
Further, the measurement model \eqref{eq:main_int} defines a relationship between the strain function and the measurements. We will use this relation to define a joint distribution between the two, which later will be used to do the inference.

First we reformulate the integrator of \eqref{eq:main_int} as
	\begin{align}
	\hat{\vec{n}}^\Transp \mat{\epsilon}(\x)\hat{\vec{n}}=	
	\begin{bmatrix}
	n_\cx & n_\cy
	\end{bmatrix}	
	\begin{bmatrix}
	\epsilon_{\cx\cx}(\x) & \epsilon_{\cx\cy}(\x) \\ \epsilon_{\cy\cx}(\x) & \epsilon_{yy}(\x)
	\end{bmatrix}	
	\begin{bmatrix}
	n_\cx \\ n_\cy
	\end{bmatrix}		
    =\underbrace{
	\begin{bmatrix}
	n_\cx^2 & 2n_\cx n_\cy & n_\cy^2
	\end{bmatrix}}_{\triangleq\oldvec{\vec{n}}^\Transp}
	\begin{bmatrix}
	\epsilon_{\cx\cx}(\x) \\
	\epsilon_{\cx\cy}(\x) \\ 
	\epsilon_{\cy\cy}(\x)
	\end{bmatrix}
	=\oldvec{\vec{n}}^\Transp\f(\x),
	\end{align}
	such that	
	\begin{equation}\label{eq:functional_li}
	\mathrm{I}(\vec{\eta})	= \vec{\vartheta}_\vec{\eta}[\f] = 
	\frac{1}{L}\int_{0}^{L} 
	\oldvec{\vec{n}}^\Transp
	\f(\x^0+s\hat{n}) ds.
	\end{equation}
	Here, $\vec{\vartheta}_\vec{\eta}$ is also considered to be an operator that maps strain functions $\f(\x)$ into LRT functions $\mathrm{I}(\vec{\eta})$.
	 This operator is also linear and as a consequence, the Gaussianity will be preserved also for the joint distribution of $\f_* = \f(\x_*)$ and $\y = [y_1, y_2 \dots, y_N]^\Transp$. We denote this joint Gaussian distribution as
	\begin{equation} \label{eq:joint}
	\begin{bmatrix}
	\y \\
	\f_*       
	\end{bmatrix}
	\sim\mathcal{N}
	\Bigg(
	\begin{bmatrix}
	\vec{0} \\
	\vec{0}
	\end{bmatrix},
	\begin{bmatrix}
	\K_\mathrm{I}+\sigma^2I & \K_*\\
	\K_*^\Transp& \K_{**}
	\end{bmatrix}
	\Bigg),
	\end{equation}	
	where $\K_{**}$ denotes the covariance of $\f_*$, $\K_\mathrm{I}+\sigma^2I$ denotes the covariance of $\y$ and
	$\K_*$ denotes the cross-covariance between $\y$ and $\f_*$.
	The covariance of $\f_*$ is provided by the covariance function from the GP prior
	\begin{align}
		\K_{**} =\E\left[\f(\x_*)\f(\x_*)^\Transp\right] = \K(\x_*,\x_*),
	\end{align}
	where we use the fact that $\E[\f(\x)]=0$.
	The cross-covariance between an LRT measurement $y_i$ and the strain function $\f_*$ can be computed based on \eqref{eq:functional_li} as
	\begin{subequations}
		\label{eq:sing_int}
	\begin{align} \label{eq:sing_int1}
		(\K_*)_{i} 
		=\E\left[\mathrm{I}(\vec{\eta}_i)\f(\x_*)^\Transp\right]
		=
		\frac{1}{L_i}\int_{0}^{L_i}
		\oldvec{\vec{n}}_i^\Transp
		\E\left[\f(\x^0_i+s\hat{n}_i)\f(\x_*)^\Transp\right] ds	 
		=
		\frac{1}{L_i}\int_{0}^{L_i}
		\oldvec{\vec{n}}_i^\Transp
		\K(\x^0_i+s\hat{n}_i,\x_*) ds,
	\end{align}	
	where $(\K_*)_{i}$ denotes the $i$th row in the matrix $\K_*$ and where we also use the fact that $\E[y_i] = \E[\mathrm{I}(\vec{\eta}_i)]+\E[\varepsilon_i]=0$. In a similar manner, we can also compute the covariance of the measurements $\E[y_iy_j] = (\K_\mathrm{I})_{ij} + \sigma^2\delta_{ij}$, where	
	\begin{equation}\label{eq:sing_int2}
	(\K_\mathrm{I})_{ij}
	 =\E[\mathrm{I}(\vec{\eta}_i)\mathrm{I}(\vec{\eta}_j)] 
	=\frac{1}{L_iL_j}
	\int_{0}^{L_j}\int_{0}^{L_i} \oldvec{\vec{n}}_i^\Transp
	\K(\x^0_i+s_i\hat{n}_i,\x^0_j+s_j\hat{n}_j)
	\oldvec{\vec{n}}_j ds_ids_j.	
	\end{equation}	
	\end{subequations}
	To specify the full joint covariance in \eqref{eq:joint}, the integrals in \eqref{eq:sing_int} can not be expected to have an analytical solution. 
	However, numerical integration can be avoided for instance by making use of the approximation technique described in Section~\ref{sec:approx}.
	
	Based on the joint distribution \eqref{eq:joint} we can condition the strain function $\f_* = \f(\x_*)$ on the measurements $\y$ to get a posterior. Due to the Gaussianity and the linear operation of conditioning, also this posterior will be Gaussian distributed according to
	\begin{subequations} \label{eq:GPreg}
	\begin{align}
	&\f_*|\y\sim\mathcal{N}\Big(\vec{\mu}_{\f_*|\y} ,\K_{\f_*|\y}\Big),
	\end{align} 
	where
		\begin{align} 
		\vec{\mu}_{\f_*|\y} & =
		\K_*^\Transp
		(\K_\mathrm{I} +\sigma^2\I)^{-1} \y,
		\label{eq:funcMean}
		\\
		\K_{\f_*|\y} & =
		\K_{**}-
		\K_*^\Transp
		(\K_\mathrm{I} +\sigma^2\I)^{-1}\K_*.		
		\end{align} 
	\end{subequations}
	The mean $\vec{\mu}_{\f_*|\y}$ is the reconstructed strain function at position $\x_*$ and its associated covariance matrix $\K_{\f_*|\y}$ encodes the uncertainty of this reconstruction.
	
	The extension to multiple test points is straightforward. 
	We then need to modify the matrices $\K_*$ and $\K_{**}$.
	Consider the set of $M$ test points $\{\x_*^j\}_{j=1}^M$.
	The dimension of $\K_*$ will change from $N\times3$ to $N\times3M$, where the columns $3j-2$ to $3j$ is built up according to \eqref{eq:sing_int1} with $\x_*=\x_*^j$.
	As for $\K_{**}$, this matrix dimension will change from $3\times3$ to $3M\times3M$, encoding the covariance between all test points.
	Hence, it will be built up by $M\times M$ blocks each of size $3\times3$, with block $(i,j)$ being $\K(\x_*^i,\x_*^j)$.
	The conditioning \eqref{eq:GPreg} is then performed in the same way. 
	This set of test points can for example be a fairly dense grid covering the whole region of interest where we want to do the reconstruction. 
	 
	 \subsection{Extension of measurement model}
	 \label{sec:GPmeasExt} 
\begin{figure}
	 		\centering
	 		\includegraphics[width=0.8\columnwidth]{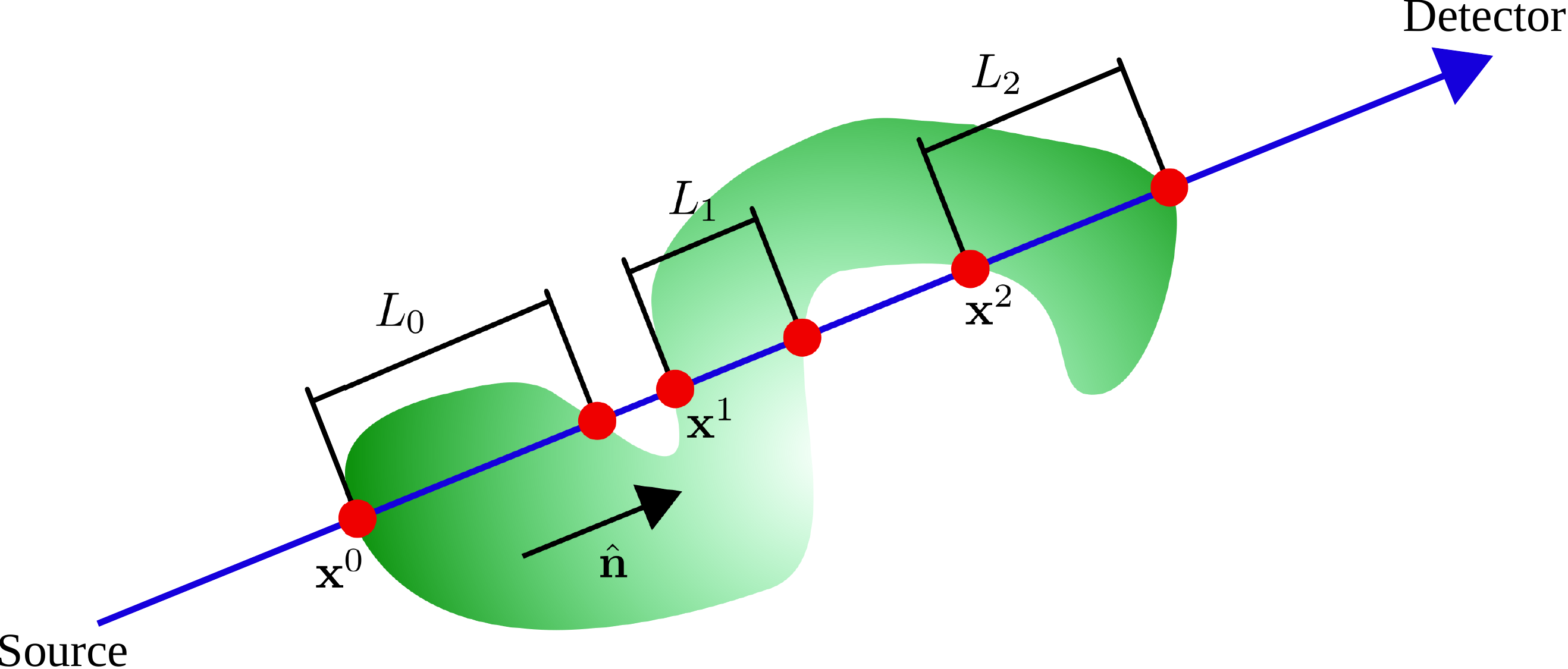}
	 		\caption{Illustration of the case where the neutron beam passes through $P=3$ segments of the sample. 
	 		}
	 		\label{fig:setup_multiSegs}
	 	\end{figure}

	 	So far we have restricted the model to the case where each neutron beam passes through the sample only once on its way from the source to the detector. 
	 	In the general case, however, we must allow the beam to pass through several segments of the sample. 
	 	We denote the starting points of the $P$ different segments with $\x^k$ and the corresponding end points with $\x^k + L^k\hat \n$ as illustrated in Fig. \ref{fig:setup_multiSegs}. 
	 	The measurement is still to be interpreted as the average strain tensor along these line segments. 
	 	Therefore, we have to integrate along all of these line segments and normalise with the total length $\sum\nolimits_{k=0}^{P-1}L^k$ travelled though the sample.
	 	The measurement equation \eqref{eq:functional_li} then turns into
	 	\begin{align}\label{eq:functional_li_gen_segs}	 	
	 	\mathrm{I}(\vec{\eta})
	 	&=\vec{\vartheta}_\vec{\eta}[\f] =
	 	\frac{1}{\sum_{k=0}^{P-1}L^k}
	 	\sum_{k=0}^{P-1}
	 	\int_{0}^{L^k} 
	 	\oldvec{\vec{n}}^\Transp\f(\x^k+s\hat{n}) ds.
	 	\end{align}
	 	Here, $\vec{\eta}$ consists of all arguments for all segments $\vec{\eta} = \{\x^0,L^0,\dots,\x^{P-1},L^{P-1},\hat \n\}$. Note that the direction $\hat \n$ is the same for all segments.
	 	Although this modification requires some care in the implementation, there is no conceptual challenge added. The covariance matrices in \eqref{eq:sing_int} will change accordingly	 	
	\begin{subequations}\label{eq:regress_matrices_multSegs}	 	
	\begin{align}
	(\K_*)_{i} 	 
	 &=\frac{1}{\sum_{k=0}^{P_i-1}L^k_i}
	\sum_{k=0}^{P_i-1}
	 	\int_{0}^{L^k_i} 
	\oldvec{\vec{n}}_i^\Transp
	\K(\x^k_i+s\hat{n}_i,\x_*) ds,\\
		(\K_\mathrm{I})_{ij}	
			 &=\frac{1}{\left(\sum_{k=0}^{P_i-1}L^k_i\right)
				\left(\sum_{l=0}^{P_j-1}L^l_j\right)} 
								\sum_{k=0}^{P_i-1}	\sum_{l=0}^{P_j-1}
				\bigg[
				\int_{0}^{L_j^l}\int_{0}^{L_i^k} \oldvec{\vec{n}}_i^\Transp
		\K(\x^k_i+s_i\hat{n}_i,\x^l_j+s_j\hat{n}_j)
		\oldvec{\vec{n}}_j ds_ids_j\bigg].	
	\end{align}		
	\end{subequations}	 			 
	The reconstruction procedure described in \eqref{eq:GPreg} remains the same with these new covariance matrices in place.

	\section{The model in practice}
	Before presenting the numerical results, we discuss a couple of important practical aspects concerning the computational complexity and the hyperparameter selection.
	The approach as described below is summarised in Algorithm~\ref{alg:batch-estimation}.

	\subsection{Reducing the computational complexity}\label{sec:approx}
		A bottleneck in GP regression is the storage and inversion of the matrix $\K_\mathrm{I} +\sigma^2\I$ in \eqref{eq:GPreg}, which scales as $\mathcal{O}(N^2)$ and $\mathcal{O}(N^3)$ in memory and time, respectively.
		For large data sets, approximate methods are motivated by the need to decrease the problem size, making memory requirements manageable and reducing the runtime.
		There are a variety of methods described in the literature, and we refer to \citep{CandelaApprox} for a useful review.
		Here we make use of the approximation method proposed by \citet{SolinSarkka}, which turns out to fit our model very well.
\modh{A drawback with this method is that it requires the covariance function to be \textit{stationary}, which means that it is dependent only on the difference between the input locations.
		However, that is true for some of the most common ones, including the squared exponential \eqref{eq:se_covfunc}.}
		 
		The key idea is to estimate the covariance function as a truncated sum of $m$ basis functions
		\begin{equation}
		k_\varphi(\vec{x,x'})\approx \sum_{i=1}^{m}S(\bs{\lambda}_i)\phi_i(\x)\phi_i(\x'),
		\end{equation}
		where $S$ is the \textit{spectral density} of the covariance function.
		For a stationary covariance function $k_\varphi=k_\varphi(\bs{r})$ where $\bs{r}=\x-\x'$, the spectral density is given by 
		\begin{equation}
			S(\bs{\omega})=\int k(\bs{r})e^{-\text{i}\bs{\omega}^\Transp\bs{r}}d\bs{r}. 
		\end{equation}
		The basis functions $\phi_i(\x)$ and eigenvalues $\bs{\lambda}_i$ are obtained from the solution to the Laplace eigenvalue problem on the domain $\Omega$
		\begin{align}\label{eq:Laplace}
		\begin{cases}
		\hspace{-4mm}
		\begin{split}
				-\Delta\phi_i(\x)&  =\|\bs{\lambda}_i\|^2\phi_i(\x),  \\
				\phi_i(\x)&         =0,         
		\end{split}
		\end{cases}
		\quad
		\begin{split}
			\x&\in\Omega, \\
			\x&\in\partial\Omega. 
		\end{split}
		\end{align}
		The Dirichlet boundary condition is the most natural choice, although any boundary condition could be chosen. 
		For a two-dimensional domain $\Omega=[-\rho_x,\rho_x]\times[-\rho_y,\rho_y]$, the solution of \eqref{eq:Laplace} reads
		\begin{subequations}
			\begin{align}
			\phi_i(\x)  &=
			\frac{1}{\sqrt{\rho_x\rho_y}}\sin\big(\lambda_{i_x}(x+\rho_x)\big)\sin\big(\lambda_{i_y}(y+\rho_y)\big) ,
			\\
			\lambda_{i_x} &= \frac{\pi i_x}{2\rho_x} ,
			\qquad \lambda_{i_y} = \frac{\pi i_y}{2\rho_y} ,                                              
			\end{align}
		\end{subequations}
		where $i_x$ and $i_y$ are chosen such that the eigenvalues lie in a desired frequency domain, and the size of $\Omega$ should be adjusted thereafter.
		Loosely speaking, the choice of $\rho_x$ determines the frequency resolution in the $x$-direction, and similarly for $y$.
		
		The approximate posterior expressions are
		\begin{subequations}\label{eq:pred_appr}
			\begin{align}
			\mathbb{E}[f_*]
			&\approx\vec{\phi}_{*}^\Transp(\mat{\Phi}\mat{\Phi}^\Transp+\sigma^2\mat{\Lambda}^{-1})^{-1}\mat{\Phi}\y,
			\\
			\mathbb{V}[f_*]
			&\approx\sigma^2\vec{\phi}_{*}^\Transp(\mat{\Phi}\mat{\Phi}^\Transp+\sigma^2\mat{\Lambda}^{-1})^{-1}\vec{\phi}_{*},
			\end{align}
		\end{subequations} 
		where $\bs{\Phi}_{ij}=\phi_i(\x_j)$, $\vec{\phi}_{*}=[\phi_1(\x_*)\dots\phi_m(\vec{x}_*)]^\Transp$ and $\bs{\Lambda}_{jj}=S(\bs{\lambda}_j)$. 
		The correct expressions for our problem are found by projecting the transformation given by \eqref{eq:functional_li} onto the basis functions.
		We end up with
		\begin{subequations}\label{eq:pred_appr_strain}
			\begin{align}
			\mathbb{E}[\f_*]
			&\approx
			\vec{Q}_{*}^\Transp(\vec{Q}\vec{Q}^\Transp+\sigma^2\mat{\Lambda}^{-1})^{-1}\vec{Q}\y,
			\\
			\mathbb{V}[\f_*]
			&\approx
			\sigma^2\vec{Q}_{*}^\Transp(\vec{Q}\vec{Q}^\Transp+\sigma^2\mat{\Lambda}^{-1})^{-1}\vec{Q}_{*}. \label{eq:pred_appr_strain_var}
			\end{align}
		\end{subequations}
		where
		\begin{subequations}\label{eq:appr_expr}
		\begin{align}
			\vec{Q}_{*} & = 
			\begin{bmatrix}
			\mathcal{L}_{\x}\phi_1|_{\x=\x_{*}} & \dots & \mathcal{L}_{\x}\phi_m|_{\x=\x_{*}}
			\end{bmatrix}^\Transp, \label{eq:basis-Lx} 
			\\
			\vec{Q}_{ij} 
			& = \vec{\vartheta}_{\vec{\eta}_j}[\mathcal{L}_\x\phi_{i}].  \label{eq:basis-thetaLx}
		\end{align}
		\end{subequations}
		Comparing this with \eqref{eq:regress_matrices_multSegs}, we have used that 
		$\K_*\approx \vec{Q}^\Transp\bs{\Lambda}\Q_*$ 
		and
		$\K_\mathrm{I}\approx \vec{Q}^\Transp\bs{\Lambda}\vec{Q}$, and the computationally more preferable form \eqref{eq:appr_expr} is obtained by utilising the identities
		\begin{align*}
			\P \B^\Transp (\B\P\B^\Transp + \R)^{-1} = (\B^\Transp \R^{-1}\B+\P^{-1})^{-1} \B^\Transp \R^{-1},
			\intertext{and}
			\A - \A \C^\Transp(\C \A \C^\Transp+\W)^{-1}\C \A
			=
			(\C^\Transp \W^{-1}\C +\A^{-1})^{-1}.
		\end{align*} 
		This approximation scheme reduces the complexity of the regression  from $\mathcal{O}(N^3)$ to $\mathcal{O}(Nm^2)$. 
		The actual savings in our case are even larger, since the hazard of numerically computing the integrals in \eqref{eq:regress_matrices_multSegs} is removed: now all we need is to compute single integrals, and this is done analytically due to the simple form of the basis functions (for details see \ref{app:approx}). 
		Hence, for this particular problem, the approximation is computationally preferable even if $N<m$.
		
		Extending to multiple test points $\{\x_*^j\}_{j=1}^M$, all we need to change is $\Q_*$, so that \eqref{eq:basis-Lx} becomes
		\begin{align}
		\vec{Q}_{*} & = 
		\begin{bmatrix}
		\mathcal{L}_{\x}\phi_1|_{\x=\x_{*}^1} & \dots & \mathcal{L}_{\x}\phi_m|_{\x=\x_{*}^1} \\
		\vdots & \vdots & \vdots \\
		\mathcal{L}_{\x}\phi_1|_{\x=\x_{*}^M} & \dots & \mathcal{L}_{\x}\phi_m|_{\x=\x_{*}^M}
		\end{bmatrix}^\Transp.
		\end{align}
		The expressions \eqref{eq:pred_appr_strain} are then applied as before.
		Note that, usually what is desired is the variance of the different component values in each test point, and not the covariance between them.
		Hence, we should not compute the entire matrix in \eqref{eq:pred_appr_strain_var}, but only its diagonal elements.  
			
		\subsection{Hyperparameters}\label{sec:hypparams}
        The covariance function $k_\varphi$ is characterised by its hyperparameters 
        $\bs{\theta}=\{\theta_k\}$. 
        An example was given in equation \eqref{eq:se_covfunc}.
	    This set does also include the noise level $\sigma$.
		Usually they are selected by maximising the marginal likelihood $p(\y|\{\bs{\eta}_i\},\bs{\theta})$, which is the probability of the data conditioned on the input locations and the hyperparameters.
		The idea is to choose the hyperparameters $\bs{\theta}_*$ that given the choice of covariance function are most likely to have generated the observed data.
		The marginal likelihood and its derivatives can be computed in closed form \citep{Rasmussen2006}.
	    For convenience, the logarithm of the marginal likelihood is usually considered, and it is for our problem given by
		\begin{align}\label{eq:ML}
		\log p(\y|\{\bs{\eta}_i\},\bs{\theta}) 
		= -\frac{1}{2}\log \det(\K_\mathrm{I}+\sigma^2I)
		-\frac{1}{2}\y^\Transp (\K_\mathrm{I}+\sigma^2I)^{-1}\y
		-\frac{N}{2}\log2\pi,
		\end{align}
		where $\K_\mathrm{I}$ is a function of $\bs{\theta}$.
		Since the constant term is irrelevant for optimisation purposes, we get
		 \begin{align}\label{eq:MLopt}
		 \bs{\theta}_*=
		 \underset{\bs{\theta}}{\text{argmax}}
		 \Big[
		 -\frac{1}{2}\log \det(\K_\mathrm{I}+\sigma^2I)
		 -\frac{1}{2}\vec{y}^\Transp (\K_\mathrm{I}+\sigma^2I)^{-1}\vec{y}
		 \Big].
		 \end{align}
		 An approximative version of \eqref{eq:MLopt} is obtained by replacing $\K_\mathrm{I}$ with $\vec{Q}^\Transp\bs{\Lambda}\vec{Q}$.  
		 The derivative expressions for this case are given in \ref{app:ML}. 
		 The optimisation can thereafter be carried out using any standard gradient-based method, such as for example the BFGS algorithm \citep{Nocedal2000numerical}.   
		
	\begin{algorithm}[t]
		\caption{Reconstructing the GP  with the reduced-rank approach}
		\label{alg:batch-estimation}
		\begin{algorithmic}[1]
			\REQUIRE $\mathcal{D} = \{(\vec{\eta}_i,y_i)\}_{i=1}^N$, $\vec{x}_*$, $\Omega$, $m$.
			\ENSURE $\mathbb{E}[\vec{f}(\vec{x}_*)], \mathbb{V}[\vec{f}(\vec{x}_*)]$.
			\STATE Construct the matrix $\Q$ as defined in \eqref{eq:basis-thetaLx}, with details given in \ref{app:approx}.
			\STATE Optimise the hyperparameters $\vec{\theta}$ as described in Section~\ref{sec:hypparams}.
			\STATE Construct $\Q_*$ as defined in \eqref{eq:basis-Lx}.
			\STATE Solve the GP regression problem by \eqref{eq:pred_appr_strain}.
		\end{algorithmic}
	\end{algorithm}

			\section{Experimental results}
		\subsection{Simulated experiment -- cantilevered rectangular plate}
		
		As a simple example illustrating the potential of the method, consider the problem presented in \citep{Wensrich2016b}.
		A cantilevered plate is subject to a vertical load at the right end, see Fig.~\ref{fig:rect_canti}. 
		The approximate equations for the strain field components are
		\begin{subequations}\label{eq:cantilev_ex_strain}
		\begin{align}
		\label{eq:cantilev_ex_strain_epsxx}
		\epsilon_{xx}&=\frac{P}{EI}(l-x)y ,
		\\[2mm] 
		\epsilon_{xy}&=-\frac{(1+\nu)P}{2EI}
		\left(\frac{h^2}{4}-y^2\right) , 
		\\[2mm]
		\epsilon_{yy}&=-\frac{\nu P}{EI}(l-x)y,		
		\end{align}
		\end{subequations}
		where $I=th^3/12$.
		Here, $I$ denotes the moment of inertia, $l$, $h$, and $t$ denotes the width, height and thickness of the plate, $P$ denotes the magnitude of the load and $E$ and $\nu$ denote Youngs modulus and Poisson's ratio, respectively.   
		We are using the same numerical values as in \citep{Wensrich2016b}, namely $E=\SI{200}{\giga\pascal}$, $\nu=0.3$, $h=\SI{10}{\mm}$, $t=\SI{6}{\mm}$, $l=\SI{20}{\mm}$ and $P=\SI{2}{\kilo\newton}$. 
		The standard deviation of the \modh{synthetic} measurement noise is here $\sigma=10^{-6}$, \modh{which was found suitable for illustration purpose}.
		The covariance function used was constructed as described in Section \ref{sec:GPmodel} with $k_\varphi(\x,\x')=\exp(-\|\x-\x'\|)$.
		
		The $\epsilon_{xx}$-component according to \eqref{eq:cantilev_ex_strain_epsxx} is illustrated in Fig. \ref{fig:cantilev_trueandmeas} together with the paths along which the line integral \eqref{eq:functional_li} have been generated.
		The remaining three figures in Fig.~\ref{fig:cantilevered} show the reconstructed component and its standard deviation using 1, 5 and 10 measurements, respectively.
		
		The reconstruction was made by building the matrices defined in \eqref{eq:sing_int}, and then applying \eqref{eq:funcMean}. 
		It is interesting to note that only 10 measurements are enough for the method to produce a reconstruction that is very hard to visually distinguish from the true function. 
		Note that the uncertainty is higher in regions further away from the measurements.
		\begin{figure}[h]
		\centering
		\resizebox{10cm}{!}{
			\begin{tikzpicture}
			\path[<->] (3,1) edge node[right] {$h$} (3,-1);
			\path[-] (2.2,1) edge (3.2,1);
			\path[-] (2.2,-1) edge (3.2,-1);
			
			\path[<->] (-1.985,1.4) edge node[above] {$l$}(2,1.4);
			\path[-] (2,1.1) edge (2,1.6);
			
			\path[->] (-2,0) edge node[above=1cm] {$y$} (-2,2);
			\path[->] (-2,0) edge node[right=5mm] {$x$} (-1,0);
			
			\node[style={draw,outer sep=0pt,thick},fill=black,fill opacity=0.05] (M) [minimum width=4cm, minimum height=2cm] {};
			
			\tikzstyle{ground}=[fill,pattern=north east lines,draw=none,minimum width=0.75cm,minimum height=0.3cm]
			\node (wall) [ground, rotate=-90, minimum width=3cm,yshift=-2.15cm] {};
			\draw (wall.north east) -- (wall.north west);
			
			\draw [-latex,ultra thick] (M.east) ++ (0.2cm,0.5cm) -- +(0,-1cm) node[above right] {$P$};
			
			\node[style={draw,outer sep=0pt,thick},fill=black,fill opacity=0.05] [xshift=4cm,minimum width=5mm, minimum height=2cm] {};
			
			\path[->] (4,0) edge node[above=1cm] {$y$} (4,2);
			\path[->] (4,0) edge node[right=5mm] {$z$} (5,0);
			
			\path[<->] (3.75,-1.2) edge node[below] {$t$} (4.25,-1.2);
			\path[-] (3.75,-1.05) edge (3.75,-1.3);
			\path[-] (4.25,-1.05) edge (4.25,-1.3);
			\end{tikzpicture}
		}
		\caption{Rectangular plate of width $l$, height $h$ and thickness $t$, cantilevered on the left side and subject to a vertical load $P$ on the right. The approximate strain components within this plate are given by \eqref{eq:cantilev_ex_strain}. }
		\label{fig:rect_canti}
		\end{figure}
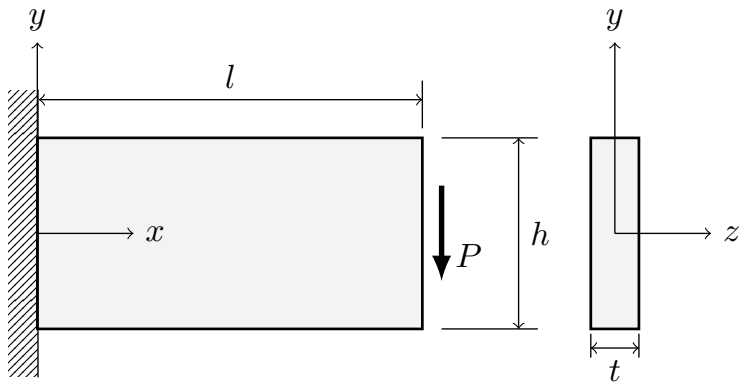
		\begin{figure}
			\centering
			\noindent\makebox[\textwidth]{
				\begin{subfigure}[t]{0.48\textwidth}
					\includegraphics[width=\textwidth]{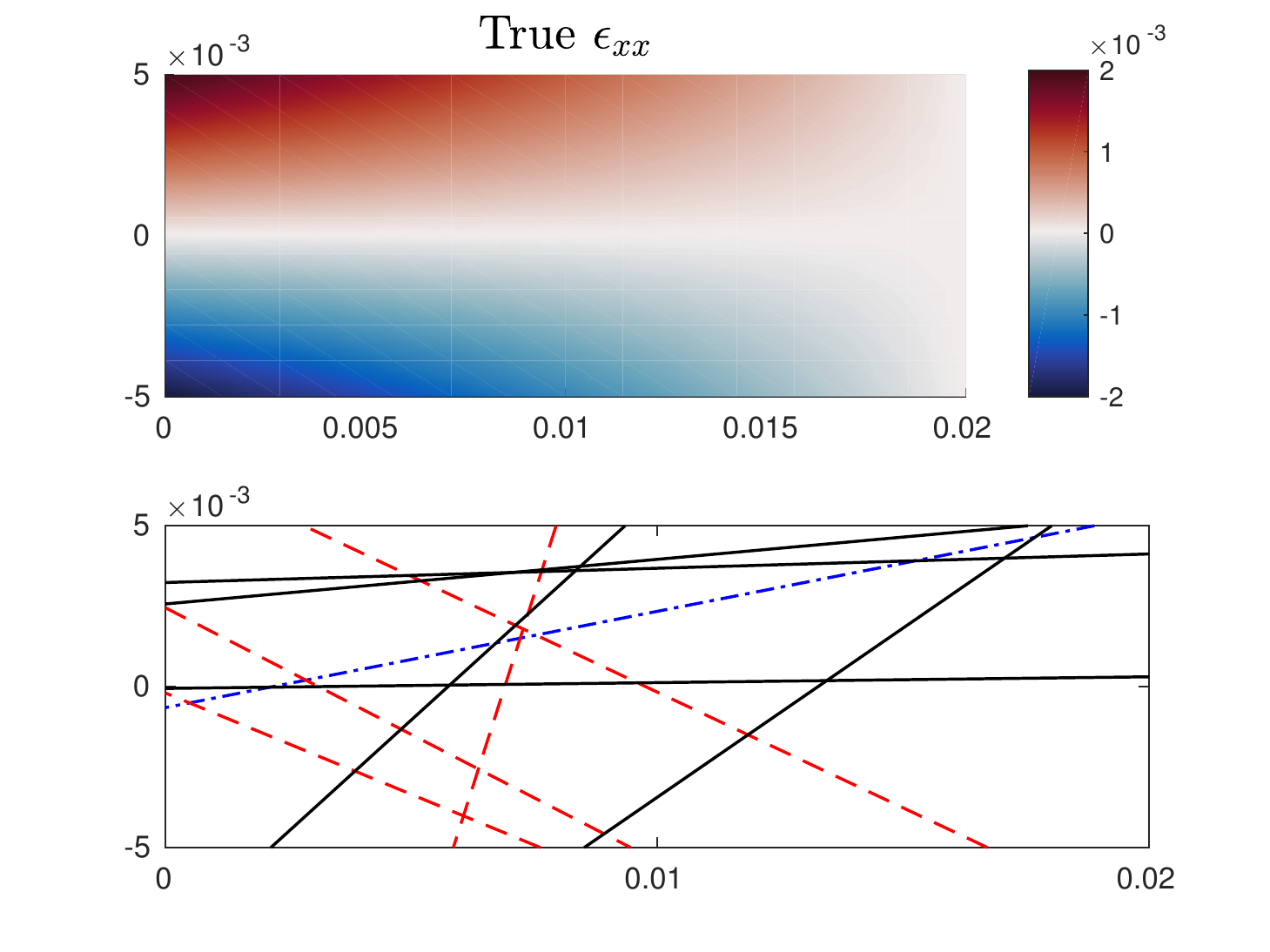}
					\caption{True component and 10 line integral measurements, number 1 (blue dashed-dot), 2-5 (red dashed), 6-10 (black solid).}
					\label{fig:cantilev_trueandmeas}
				\end{subfigure}
				\hspace{3mm}
				\begin{subfigure}[t]{0.48\textwidth}
					\includegraphics[width=\textwidth]{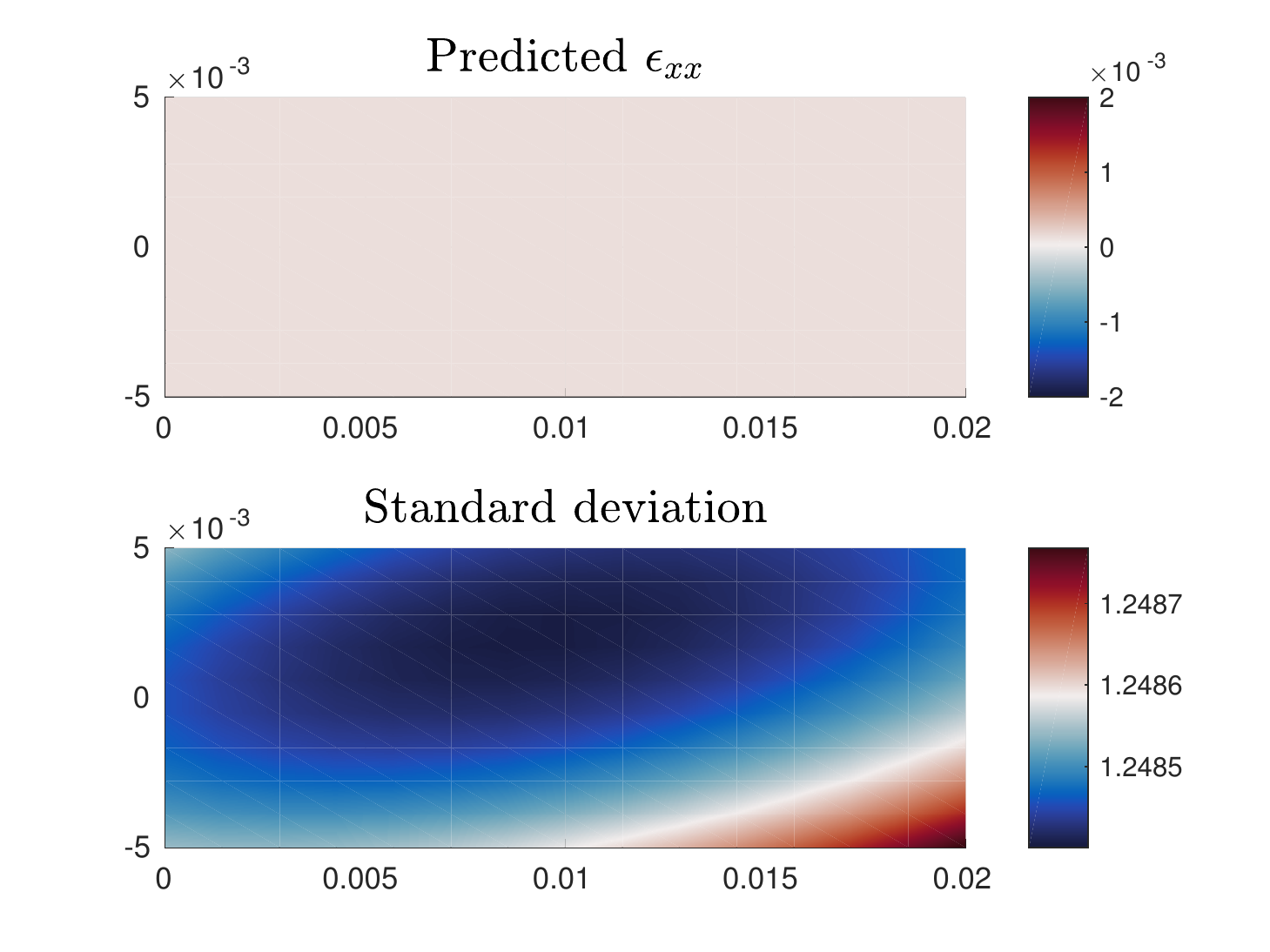}
					\caption{Prediction (top) and its standard deviation (bottom) after the first measurement (blue dashed-dot line).}
					\label{fig:cantilev_epsxx_1}
				\end{subfigure}
			}
			\noindent\makebox[\textwidth]{
				\hspace{-0mm}
				\begin{subfigure}[t]{0.48\textwidth}
					\includegraphics[width=\textwidth]{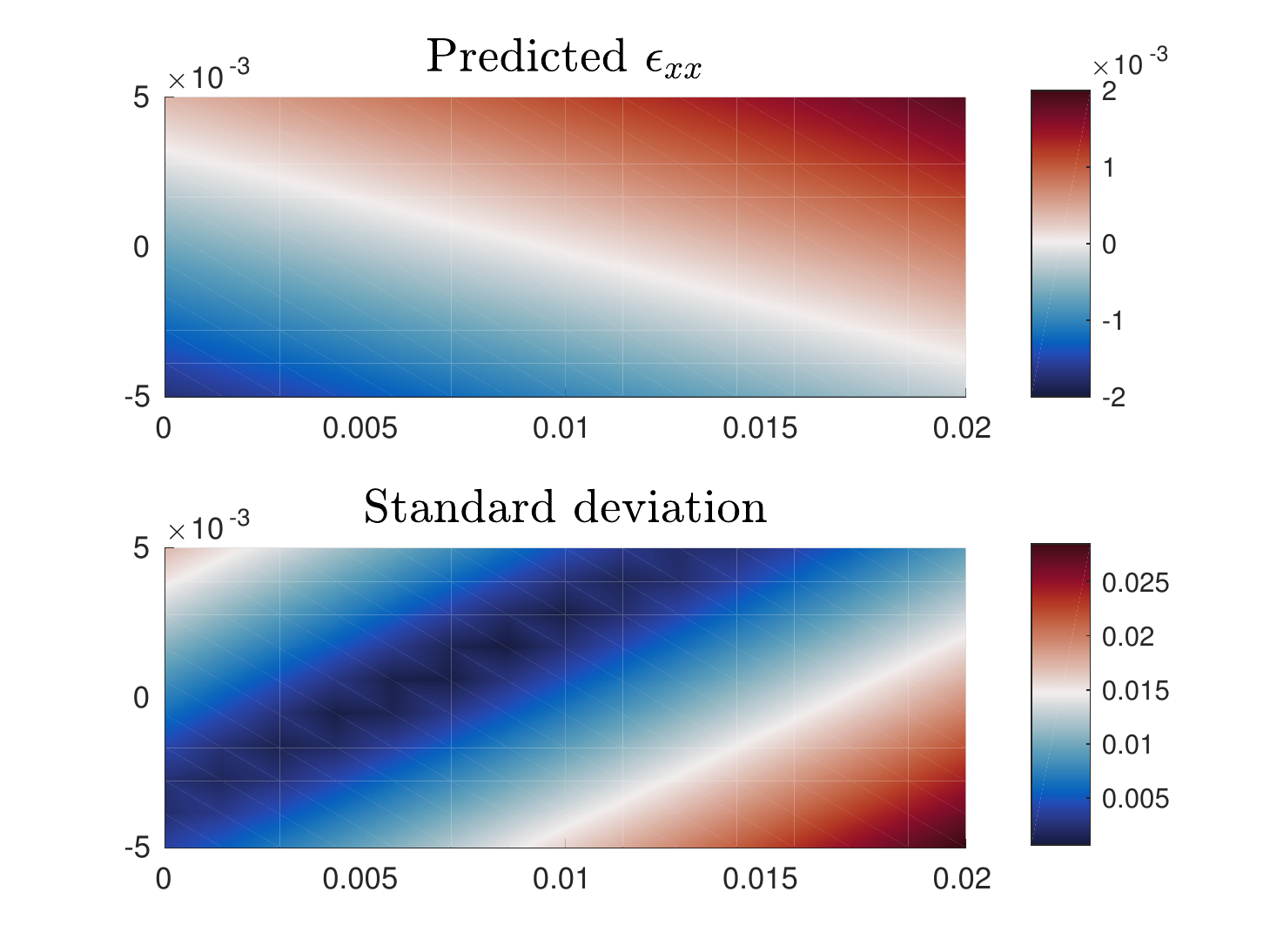}
					\caption{Prediction (top) and its standard deviation (bottom) after 5 measurements (blue dashed-dot line and red dashed lines).}
					\label{fig:cantilev_epxx_2}
				\end{subfigure}
				\hspace{3mm}
				\begin{subfigure}[t]{0.48\textwidth}
					\includegraphics[width=\textwidth]{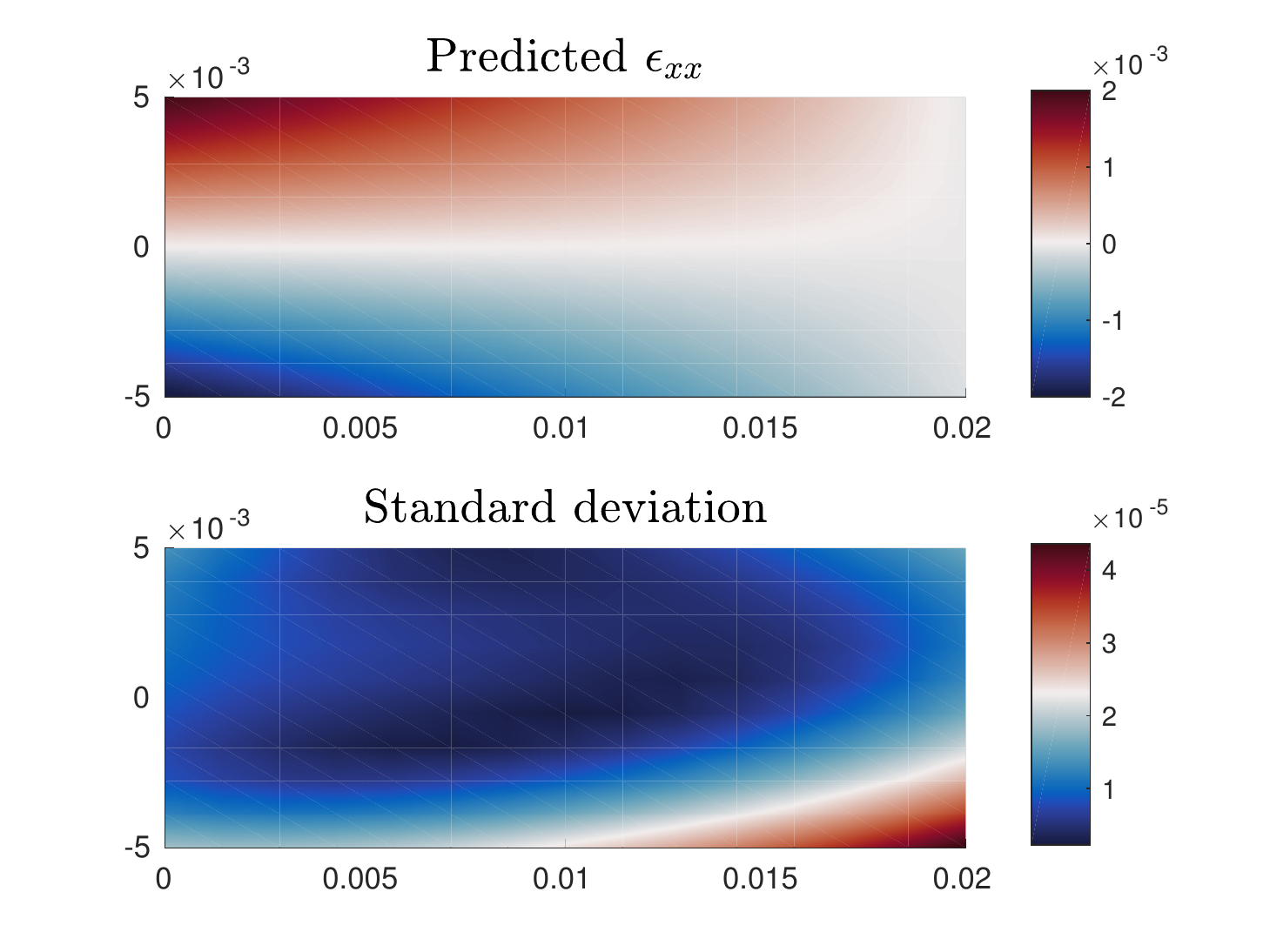}
					\caption{Prediction (top) and its standard deviation (bottom) after all measurements.}
					\label{fig:cantilev_epsxx_3}
				\end{subfigure}
			}
			\caption{True and reconstructed strain field including its standard deviation for the $\epsilon_{xx}$-component in the rectangular plate shown in Fig.~\ref{fig:rect_canti}. The measurements are the line integral paths shown in the top left figure. Note that the uncertainty is larger at the regions that are poorly covered by the measurements.}
			\label{fig:cantilevered}
		\end{figure}

		For comparison, we are performing a convergence study similar to the one performed by \citet{Wensrich2016b}.
		Here, the measured data is not generated using \eqref{eq:cantilev_ex_strain} but instead from a finite element solution of the problem, which should better reflect reality.	
		To line up with the real experimental setting, the data is not randomly chosen but comes in chunks corresponding to different \textit{projections}.
		Each projection typically contains a couple of hundred measurements taken from the same angle but uniformly distributed along the perpendicular direction. 
		
		We consider the \modh{synthetic} noise level $\sigma=10^{-4}$, \modh{to align with \citet{Wensrich2016b} where that value was used as an expected experimental measurement noise level}.
		To model the Airy stress function, we are here using the Mat\'{e}rn$_{5/2}$ covariance function
		\begin{subequations}
			\begin{align}\label{eq:matern52_2d}
			k_\varphi(\x,\x')&=
			\sigma_f^2
			\left(
			1+\sqrt{5}\tilde{r}+\frac{5}{3}\tilde{r}^2
			\right)
			e^{-\sqrt{5}\tilde{r}}, \\
			\tilde{r}^2 &= l_x^{-2}(x-x')^2+l_y^{-2}(y-y')^2,
			\end{align}
		\end{subequations}
		with the hyperparameters chosen by maximising the marginal likelihood \eqref{eq:MLopt}.
\modh{This covariance function belongs to a generalisation that relaxes the extreme smoothness assumptions of the squared exponential covariance function \eqref{eq:se_covfunc}, and is often considered to be somewhat more realistic.}
		The prediction is made in uniformly distributed points on a $40\times20$ mesh.
		
		We are reporting the relative error in the reconstruction, where this involves a concatenation of all components in all points.
		For a total set of $N$ projections, the angle from which projection $k$ is taken has been chosen as 
		$\frac{\pi}{96}+\ceil{95\frac{k-1}{N-1}-0.5}\frac{\pi}{96}$, so the projection angles are approximately evenly spaced over $[0, \,\, \pi]$.
		Gaussian noise has been added to the measurements with \textsc{Matlab}:s default random seed.
		
		We have here used the approximative method described in Section \ref{sec:approx} with $\rho_x=3l$, $\rho_y=3h/2$ and a total number of 160 basis functions. 
		The spectral density of the Mat\'{e}rn$_{5/2}$ covariance function \eqref{eq:matern52_2d} is given by
		\begin{equation}\label{eq:matern52_SD}
		S(\bs{\omega})=
			\sigma_f^2
			100\sqrt{5}\pi l_x l_y
			(5+l_x^{2}\omega_x^2+l_y^{2}\omega_y^2)^{-7/2}
		\frac{\Gamma(7/2)}{\Gamma(5/2)},
		\end{equation}
		where $\Gamma(\cdot)$ is the gamma function.
		
\begin{figure}
			\centering
			\includegraphics[width=1\textwidth]{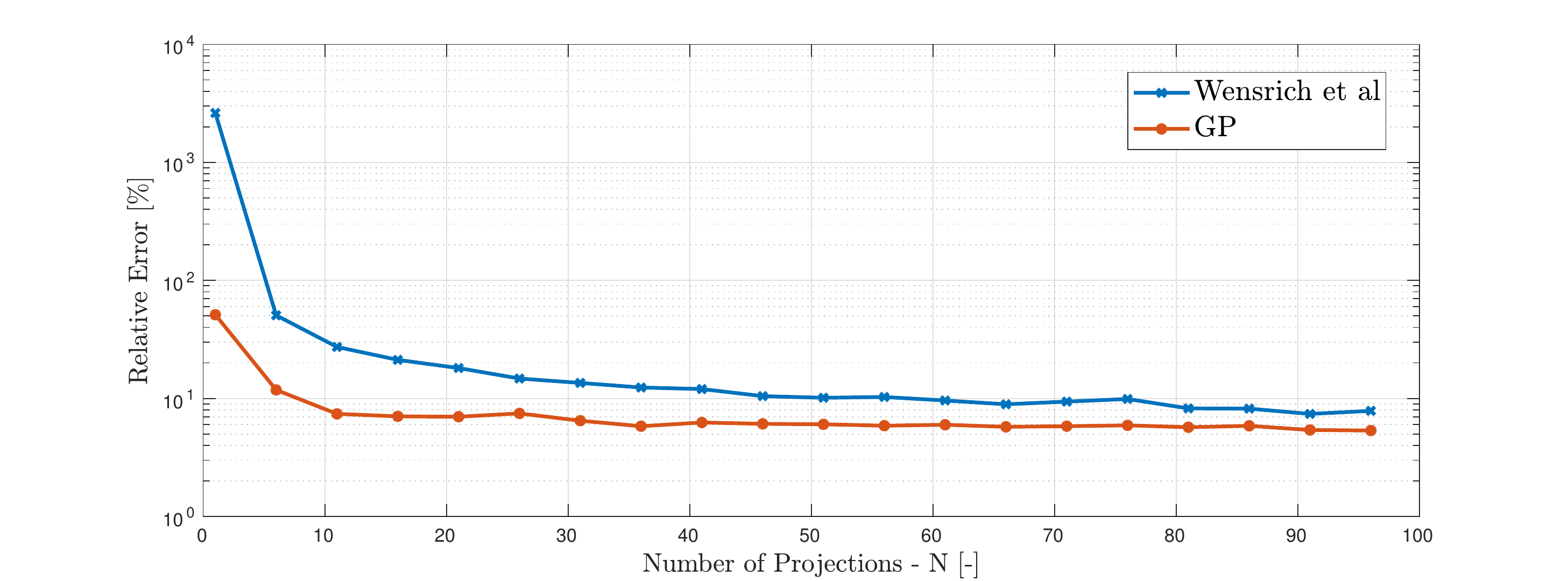}
			\caption{Relative error in the GP-reconstruction as a function of the number of projections.}
			\label{fig:cantilevered_convergence_linear}
		\end{figure}

		The result is shown in Fig.~\ref{fig:cantilevered_convergence_linear}, which also contains the corresponding curve from \citep{Wensrich2016b} for comparison.
	    It is seen that the GP-reconstruction is more accurate and requires less measurements to achieve an equivalent performance.  
		
		\subsection{Real data}\label{sec:real-data}
		In this section we are solving the problem using data collected from a real-world experiment at the Japan Proton Research Accelerator Complex (JPARC).
		A brief description of the experimental settings is given below, and we refer to \citet{Hendriks2017} for details.
		
		The sample considered is a thin C-shaped steel plate subject to a compressive load of roughly $\SI{7}{kN}$. 
		In polar coordinates it is defined by
		$r_{\text{in}}=\SI{3.5}{mm}<r<r_{\text{out}}=\SI{10}{mm}$ 
		and 
		$\SI{45}{\degree}<\theta<\SI{315}{\degree}$. 
		
		The data set consists of 86 projections taken from evenly distributed angles around the sample.
		\modh{The experimental resolution of the detector is 512x512. Since this particular experiment was dealing with a planar strain field the detector pixels could be grouped by column (in the out of plane direction) to improve the statistics, giving a possible 512 measurements per projection. 
		However, due to the sample size and shape not all rays would have passed through the sample.
		On average roughly 350 measurements were made per projection, giving a total amount of nearly 30 000 measurements. 
		It should be noted that the sample took up slightly less than half the detectors height (as another sample was also being analysed) and so about 200 pixels were binned for each measurement.}
     		
		The Mat\'{e}rn$_{5/2}$ covariance function \eqref{eq:matern52_2d} was used, and for the approximate settings we have taken $\rho_x=\rho_y=2.5r_ {\text{out}}$ and a total number of 673 basis functions.
		
\begin{figure}
			\centering
			\includegraphics[width=0.9\textwidth]{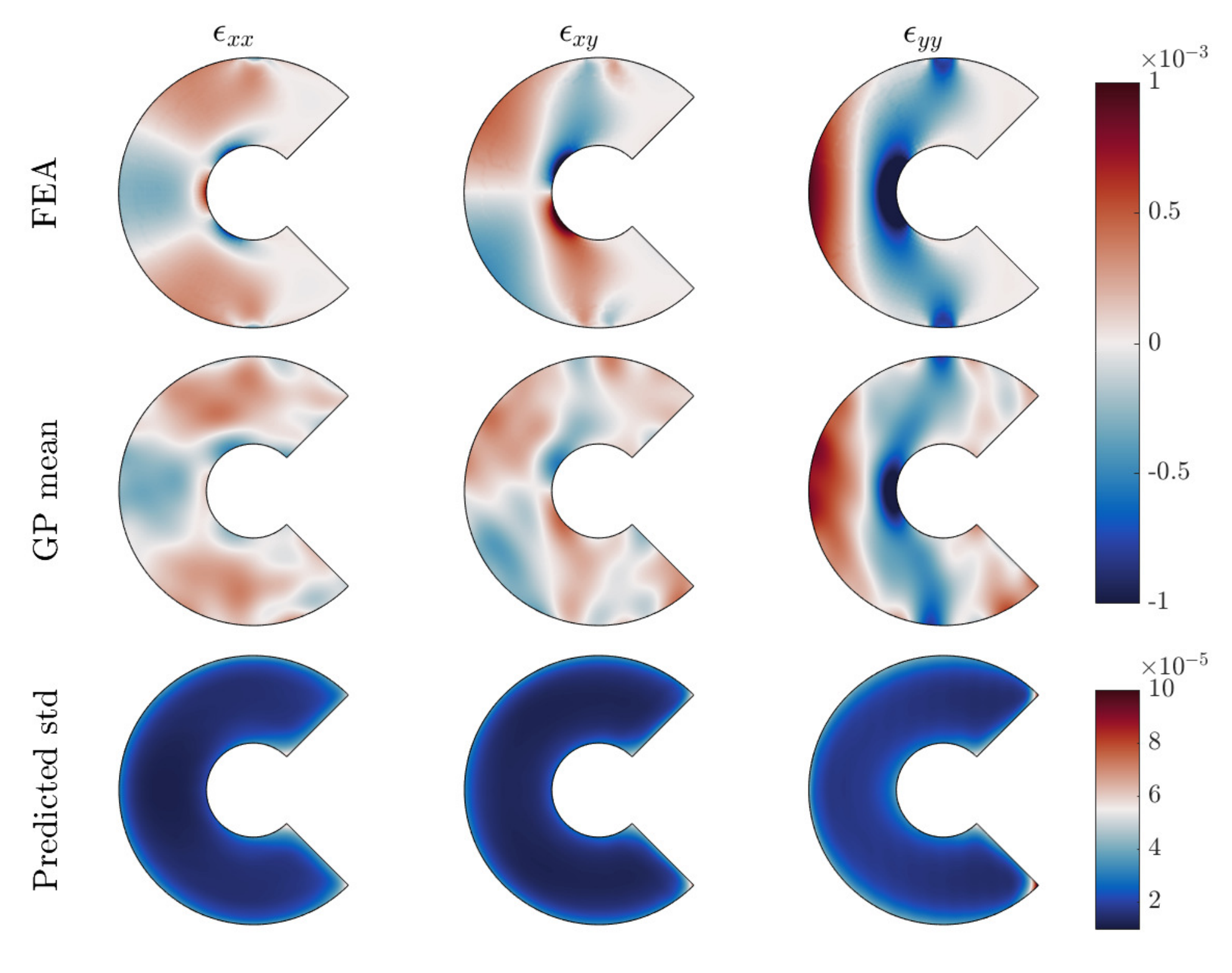}
			\caption{Top row: simulated solution obtained through finite element analysis. Middle and bottom row: mean and standard deviation of the GP reconstruction.}
			\label{fig:real-data_pcolors}
		\end{figure}

\begin{figure}
			\centering
			\includegraphics[width=\textwidth]{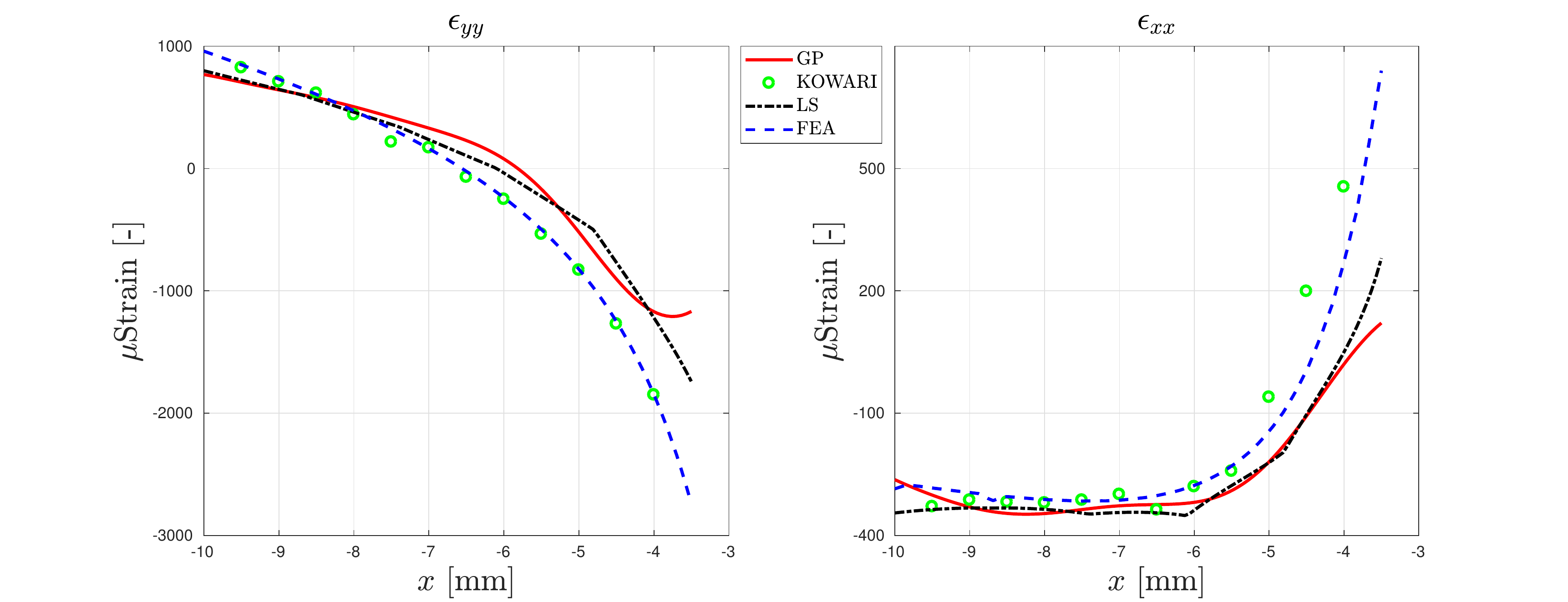}
			\caption{Plots of the GP reconstruction along the line $y=0$, compared to constant wavelength scanning (KOWARI), a finite element simulation (FEA) and a reconstruction obtained through the least squares approach (LS).}
			\label{fig:real-data_slices}
		\end{figure}

		The result is shown in Fig.~\ref{fig:real-data_pcolors}.
		The top row shows a finite element solution to the simulated problem, while the middle and bottom rows contains the mean and standard deviation of the GP reconstruction.
		Fig.~\ref{fig:real-data_slices} is showing the $\epsilon_{yy}$- and $\epsilon_{xx}$-components along the line $y=0$.
		The data points labelled KOWARI emerges from a constant wavelength scanning experiment, which is a well established method for accurate measurement of average strain within a gauge volume \citep{Hendriks2017}.
		Also, LS denote the reconstruction obtained from the least squares approach used by \citet{Hendriks2017}. 
		
		Although the GP reconstruction follows the overall structure of the FEA solution and the KOWARI data, there are regions of notable deviations. 
		This is particularly clear at the right sides in Figure \ref{fig:real-data_slices}, near the boundary of the sample. 
		It has been observed that this deviating behaviour is present when the reconstruction is made from simulated data as well.
		The boundary challenges are intuitively understood from the nature of the model.
		From the GP's perspective, the inferred function is a continuous object, and the natural problem boundary of the sample is not built into the model.
		Outside the sample the reconstruction will fall back to the prior mean, which obviously has a higher impact on the boundary than on the interior.
		This effect gives rise to a perceived non-smooth feature, which is hard to capture with the relatively smooth covariance function provided by Mat\'{e}rn$_{5/2}$.
		
		Moreover, it is not obvious what settings to choose for the approximation method. 
		In theory, the approximation improves as the number of basis functions is increased.
		In practice, too many basis functions entails numerical problems, while too few gives a poor approximation.
		This trade-off requires a somehow ad hoc user selection and the precise impact on the solution is hard to anticipate.
		
		Data-specific error sources related to the collection and processing of the raw data are discussed more thoroughly by \citet{Hendriks2017}.
		For example, certain ray paths are short relative to others, and the Bragg-edge estimate hence becomes less accurate in these cases.
		See \citet{Vogel2000} for more discussions on systematic error sources in this context. 
		
		\section{Concluding remarks}
		
		
		
		\subsection{Compatibility constraint}\label{sec:comp}
		If the strain field can be expressed as emerging from a continuous, single-valued displacement field, then it is called \textit{compatible}. 
		Compatibility can be expressed as the linear constraint
		\begin{equation*}
		\underbrace{
			\begin{bmatrix}
			\frac{\partial^2}{\partial y^2} & -\frac{\partial^2}{\partial x\partial y} & \frac{\partial^2}{\partial x^2}
			\end{bmatrix}
		}_{\Fcalmat_\x^C}
		\f
		=0.
		\end{equation*}  
		If we would like to construct a model with this constraint imposed as well, the scalar function $\varphi(\x)$ can no longer be chosen arbitrary. 
		Instead it will be governed by the relation
		\begin{equation}
		\Fcalmat_\x^C\mathcal{L}_\x \varphi=0 ,
		\end{equation}
		where $\mathcal{L}_\x$ is given by \eqref{eq:calL1}.
		This is the strain formulation of the \textit{biharmonic equation}, with solutions called \textit{biharmonic functions}. 
		As shown by \citet{Sadd}, separation of variables can be used to express a general solution $\varphi^*$, where the specific form is determined by the boundary conditions of the problem, and so is the number of parameters. 
		
		This gives rise to at least two questions. 
		Firstly, if there is a parametric form of the solution, why would we use a non-parametric regression method? 
		The answer is that a parametric model require the exact form of the solution so that the number of parameters to be estimated is known. 
		This can only be done provided accurate information of the boundary conditions, which in general can not be assumed to be available. 
		Furthermore, for some boundary conditions the analytical solution may have to be expressed as an infinite sum, thus involving an infinite number of parameters, which makes a parametric model unsuitable.  
		
		The second question is, can we include the knowledge of the general solution in the GP model? 
		In theory, the answer is yes. 
		Let $\varphi^*=\varphi_{\w}^*$ denote that the function is parametrised with the parameters in the vector $\w$. 
		Treating these parameters as random variables and assigning them the distribution $p(\w)$ allows us to calculate the covariance function of $\varphi_\w^*$
		\begin{equation}\label{eq:cov_zero_mean}
		k_{\varphi^*_\w}(\vec{x,x'})=
		\int \varphi_{\w}^*(\x)\varphi_{\w}^*(\x') p(\w)d\w.
		\end{equation}
		To be general, we must assume $\w$ to be infinite-dimensional. 
		This fact itself does not imply that an analytical expression could not be found, but the specific form that $\varphi^*$ requires may not allow it. 
		In any case, we are leaving this as a potential topic of further research.

	\subsection{Samples with grain texture}
		\modh{The GP prior was designed such that any strain fields generated would automatically satisfy the equilibrium constraints as written in \eqref{eq:equi_constr_stress1}. 
		These equilibrium equations are valid for planar, isotropic, linearly elastic samples under the assumption of plane stress. 
		Although the framework presented for strain tomography using GPs does not explicitly prevent the adaptation of this model to allow for samples with strong grain texture, there are significant challenges.
		These challenges are present both in the acquisition of transmission strain measurements and in the reconstruction from these measurements.
		During measurement acquisition strong texture may result in the Bragg-edge of interest being unobservable from particular measurement directions, however current research into full pattern fitting may provide a solution to this problem in the future \citep{sato2017further,sato2013upgrade,sato2011rietveld}. 
		During reconstruction, texture is problematic as different strain measurements may relate to the material bulk strain in different ways. If this relationship can be quantified it may be possible to build it into the GP model.}

	\subsection{Relation to diffraction measurements}
		\modh{Since comparison with a diffraction-based method (KOWARI) was made in Section \ref{sec:real-data}, we would like  to add a brief comment on the relation between these methods.  
		We would argue that tomographic methods such as the one presented in our paper and diffraction measurements are complementary rather than opposing.
	
		It is important to recognise the differences between these methods. 
		Diffraction-based methods provide measurements of average strain within a gauge volume at defined locations, whereas tomographic methods provide the full strain field over the sample.
		In addition, future increases in beam power at JPARC and other facilities should reduce the time required to collect Bragg-edge transmission measurements.
		Under these conditions, one could expect to see a significant reduction in time required to determine the full strain field
		
		
			
		
		Therefore we would suggest that if a particular area of interest is known, diffraction measurements may be a good choice, whereas if it is not known and the user would like to analyse the full field the method presented in this paper may be preferable.}
		
	\subsection{Future work}
		First of all, recall that this paper only considers the two-dimensional problem.
		Since the real world is three-dimensional, it is natural to extend the method accordingly.
		The main difference is that the target function \eqref{eq:target_func} becomes six-dimensional since a three-dimensional strain field has six unique components.
		Following this, the equilibrium constraints take an extended form, which obviously is reflected in the covariance model.
		However, the challenges are most likely not conceptual, but rather limited to the implementation. 
		
		Another interesting topic relates to the discussion in Section \ref{sec:comp}, where we introduced the idea of constructing a covariance function for compatible strain fields based on the theoretical solution.
		Although this appears to be intractable and perhaps not desirable, we did not investigate the subject any further.
		However, there may be situations in which a theoretical formulation can be derived in an exact or approximate form when this idea may be applicable -- especially if the strain field is governed by relatively simple equations.

		\modh{Furthermore, one may want to explore other covariance models.
		As always, simple alternatives such as the squared exponential and the Mat\'{e}rn covariance functions should always be tried first, since they have shown to perform well in many applications.
		As for the particular problems we have considered, there has not been enough indications motivating the implementation of a more advanced option.
		Also, non-stationary alternatives as discussed below can not be used with the approximation method employed in this paper.}
		
		However, it may be situations where extensions are necessary.
		An example is if the inferred function contains significant non-smooth features, such as rapid changes or discontinuities.
		In those cases it may be hard to obtain a satisfying reconstruction with a simple model -- this is referred to as \textit{mis-specification}.
		As shown by \citet{Rasmussen2006}, such situations can be dealt with by building the covariance function as a sum of several terms encoding different properties, or changing to another covariance function that is better suited to the data. 
		Even more powerful models having gained interest in recent years are the so-called \textit{deep GPs} \citep{DamianouPHD} and the related \textit{manifold GPs} \citep{Calandra2014}.
		
		Particularly, in some situations the strain behaviour varies significantly between different well-specified parts of the domain, although nearby located.
		An example of this is the ball bearing problem illustrated by \citet{Wensrich2016a}.
		A way to deal with this might be to use different GPs in each subdomain, conceptually similar to the piecewise GP approach employed by \citet{Svensson2017189} or a so-called \textit{mixtures of experts} model \citep{Tresp2001}. 
		The most challenging part here is to find a neat way of making use of data spanning multiple subdomains, as is the case with line integral measurements.
		An extended covariance model as outlined above is a potential alternative for this problem as well. 
		\modh{For detailed discussions on the model selection problem, see e.g. \citet{Rasmussen2006}.}

		
	
		\subsection{Conclusion}
		In this paper we have introduced the concept of probabilistic modelling within the field of tomographic reconstruction.
		In particular, we have shown that Gaussian processes can be used for strain field estimation from Bragg-edge measurements.
		The probabilistic nature of the model allows for a systematic treatment of the noise and it provides a direct uncertainty measure of the reconstruction.
		We have shown that known physical laws can be explicitly incorporated in the design of the associated covariance function, relying on the property that Gaussian processes are closed under linear transformations.   	
		Experiments performed on simulated and real data indicates that the method has a high potential which opens up for other tomographic applications as well. 

\section{Acknowledgements}
	This research was financially supported by the Swedish Foundation for Strategic Research (SSF) via     the project \emph{ASSEMBLE} (contract number: RIT15-0012).

\appendix
	\section{Bragg-edge method}\label{sec:BragEdgeMethod}
	A well-established method for strain estimation within deformed polycrystalline materials relies upon so-called Bragg-edge analysis \citep{Santisteban_Eng}.
	A summary of the procedure goes as follows.
	
	The sample investigated is penetrated by neutron beams in a two-step procedure -- before and after the deformation occurs.
	The neutron beams contains a spectra of wavelengths, and they are transmitted in pulses each with a well-known relation between wavelength and intensity.
	After having passed through the material, the intensity of the beams is recorded at a detector.
	The wavelength profiles can be measured because of the direct relationship between velocity and wavelength and hence the recorded arrival time of the neutron at the detector is a proxy for the wavelength.
	
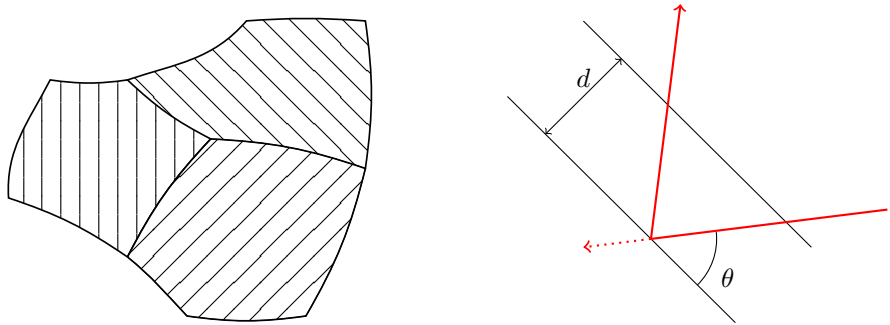
\begin{figure}
			\centering
			\resizebox{5cm}{!}{
				\begin{tikzpicture} 	
				\draw [fill=white,pattern=custom vertical lines,hatchspread=7.07pt,fill opacity=1,thick,draw=black,draw opacity=1]
				(0,0) to[bend right=8mm] (-2,1) to[bend left=8mm] (-1.3,3) to[bend right=8mm] (0,3) to[bend left=8mm] (1.4,2) to[bend right=8mm] (0,0);
				
				\draw [fill=white,fill opacity=1,draw opacity=0] 
				(4,1.5) to[bend right=8mm] (4,4) to[bend right=4mm] (2,4) to[bend left=8mm] (0,3) to[bend right=8mm] (1.4,2);
				\draw [pattern=custom north west lines,hatchspread=10pt,thick,draw=black,draw opacity=1] 
				(4,1.5) to[bend right=8mm] (4,4) to[bend right=4mm] (2,4) to[bend left=8mm] (0,3) to[bend right=8mm] (1.4,2);

				\draw [fill=white,fill opacity=1,draw=black,thick,draw opacity=1]
				(0,0) to[bend left=2mm] (1,-1) to[bend right=8mm] (3,-1) to[bend right=8mm] (4,1.5) to[bend right=8mm] (1.4,2) to[bend right=8mm] (0,0);
				\draw [pattern=custom north east lines,hatchspread=10pt]
				(0,0) to[bend left=2mm] (1,-1) to[bend right=8mm] (3,-1) to[bend right=8mm] (4,1.5) to[bend right=8mm] (1.4,2) to[bend right=8mm] (0,0);
				\end{tikzpicture}
			}
			\hspace{15mm}
			\begin{tikzpicture} 	
			\path[-] (-1,4) edge (2,1); 
			\path[-] (0,5) edge (3,2); 
			\path[red,-] [thick] (4,2.5) edge (0.8889,2.1111);
			\path[red,->] [thick] (0.8889,2.1111) edge (1.27778,5.2222);
			\path[red,dotted,->] [thick] (0.8889,2.1111) edge (0,2);
			\path[<->] (-0.5,3.5) edge node[above] {$d$} (0.5,4.5);
			\draw[-] (1.5,1.5) arc (-45:6:0.86424);
			\draw node at (1.9,1.6) {$\theta$};
			\end{tikzpicture}
			\caption{\textit{Left}: Simple illustration of a crystal structure. The material is built up by crystals with planes of certain lattice spacing directed in different angles. 
				\textit{Right}: A ray incident on the material will interact with the crystals whose planes are directed in an angle such that Bragg's law \eqref{eq:braggslaw} is fulfilled.}
			\label{fig:crystal}
		\end{figure}

	The material contains a very large number of randomly oriented crystal planes that the neutrons interact with, and constructive diffraction occurs according to \textit{Bragg's law}
	\begin{equation}\label{eq:braggslaw}
		\lambda=2d\sin\theta,
	\end{equation} 
	where $\lambda$ is the neutron wavelength, $d$ is the lattice spacing between the crystal planes and $\theta$ is the scattering angle, see Fig.~\ref{fig:crystal}.
	The neutrons are scattered up until $\theta=90^\circ$, a point at which they are reflected back towards the incoming direction -- so-called \textit{backscattering}.
	For larger wavelengths, no scattering can occur which results in a sudden increase in the relative transmission rate \modh{(the ratio of the open beam intensity when no sample is present and the measured intensity when the sample is present)}.
	This is known as a \textit{Bragg-edge}. 
	The change in position of the Bragg-edges due to the deformation of the sample is used to calculate a measure of the average strain $\langle\epsilon\rangle$ along the propagating direction of the neutron beam
	\begin{equation}\label{eq:av_strain}
		\langle\epsilon\rangle = \frac{d-d_0}{d_0},
	\end{equation}
	where $d_0$ and $d$ denote the lattice spacings before and after deformation, respectively.
	Since the material consists of a large number of lattice spacings, each measurement contains several Bragg-edges. 
	In practice a Bragg-edge is chosen that is characteristic of the materials bulk properties (elastic modulus) and also has a good Bragg-edge height (dependent on the source spectra etc).
	A measurement of the form \eqref{eq:av_strain} is modelled with the LRT \eqref{eq:main_int}.	

	\section{Gaussian processes under linear transformations}\label{sec:func_observs}
	A useful property of the GP is that it is closed under linear functional transformations \citep{Papoulis1991,Rasmussen2006,Hennig2013,Garnett2017,WahlstromPHD}. 
	This means that if  
	\begin{equation*}
	f(\x)\sim\GP\big(m(\x),k(\vec{x,x'})\big),
	\end{equation*}  
	then 
	\begin{equation*} 				
	L_{\w}[f(\x)]
	\sim\GP\Big(L_{\w}[m(\x)],L_{\vec{w,w'}}^2[k(\vec{x,x'})]\Big) ,
	\end{equation*}  
	where $L_{\w}$ is a linear functional with argument $\w$, and $L^2_{\w,\w'}$ indicates that it is acting on both arguments of $k$. 
	With $L_{\w}$ being linear we mean that
	\begin{equation}
	L_{\w}[\alpha f(\x) + \beta g(\x)]=
	\alpha L_{\w}[f(\x)] + \beta L_{\w}[g(\x)],
	\end{equation}
	for the two scalars $\alpha$ and $\beta$.
	Two common and important examples of linear functionals are differentiation 
	\begin{align}
	\mathcal{D}_{i,\vec{\zeta}}[f]&=\pd{f}{x_i}\Big|_{\x=\vec{\zeta}} ,
	\intertext{and integration} 
	\mathcal{I}_\Omega[f]&=\int_\Omega f(\x)d\x.
	\end{align}
	The key here, which makes the closure property so useful, is that the function and the functional have a joint Gaussian distribution. 
	This implies that predictions of the function can be conditioned on observations of the functional, and vice versa. 
	This property is useful in regression problems where we can not observe the function directly.
	
	For example, consider integration of a one-dimensional function $f(x)$ over the interval $\Omega=[a \,\, b]$. 
	If we model the function with a GP
	\begin{equation}
	f(x)\sim
	\GP(m(x),k(x,x')),
	\end{equation}
	then it follows from the above that
	\begin{align} 
	L_{\w=\Omega}[f]
	=\mathcal{I}_\Omega[f]=z(\Omega)=
	\int_a^b f(x)dx
	\sim
	\GP\left(\int_a^b\mu(x)dx,\int_{a'}^{b'} \int_{a}^{b} k(x,x')dx dx'\right). 
	\end{align}  
	Note that the input to this GP is not the variable $x$, but the parameterisation of the integration interval $\Omega$.
	More concretely, assume that we want to predict the value of $f(x_*)$ from integral measurements of $f(x)$. 
	An element in the Gram matrix then becomes 
	\begin{align}
	\K_{ij} =\int_{a_j}^{b_j}\int_{a_i}^{b_i}k(x,x')dxdx', 
	\end{align} 
	which describes the correlation between 
	$\int_{a_i}^{b_i}f(x)dx$ 
	and 
	$\int_{a_j}^{b_j}f(x')dx'$
	, respectively. 
	We then build the vector $\k_*$ according to
	\begin{equation}
	(\k_*)_i =\int_{a_i}^{b_i}k(x_*,x')dx',
	\end{equation} 
	which is the correlation between
	$\int_{a_i}^{b_i}f(x)dx$ 
	and the function value $f(x_*)$.
	The GP regression is performed as usual
		\begin{subequations}
			\begin{align} 
			\mathbb{E}[{f_*|\y}] & =
			\k_*^\Transp
			(\K +\sigma^2\I)^{-1} \y,                          
			\\
			\mathbb{V}[{f_*|\y}] & =
			k(x_*,x_*)-
			\k_*^\Transp
			(\K +\sigma^2\I)^{-1}\k_*.		
			\end{align} 
		\end{subequations}
	
\begin{figure}
		\centering
		\noindent\makebox[\textwidth]{
			\begin{subfigure}[t]{0.48\textwidth}
				\includegraphics[width=\textwidth]{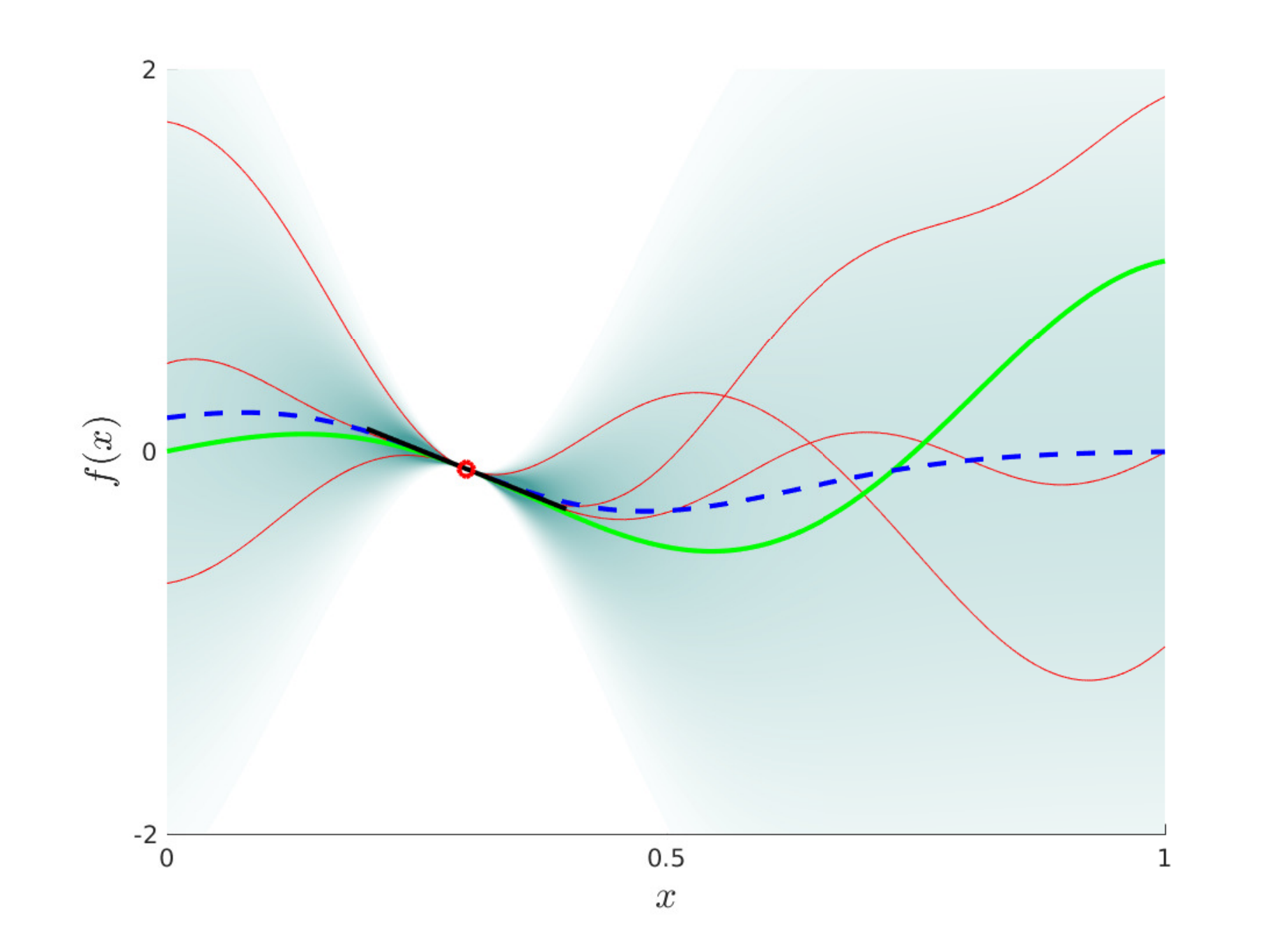}
				\caption{Posterior after one function observation and one derivative observation.}
				\label{fig:funcobs_der}
			\end{subfigure}	
			\hspace{4mm}
			\begin{subfigure}[t]{0.48\textwidth}
				\includegraphics[width=\textwidth]{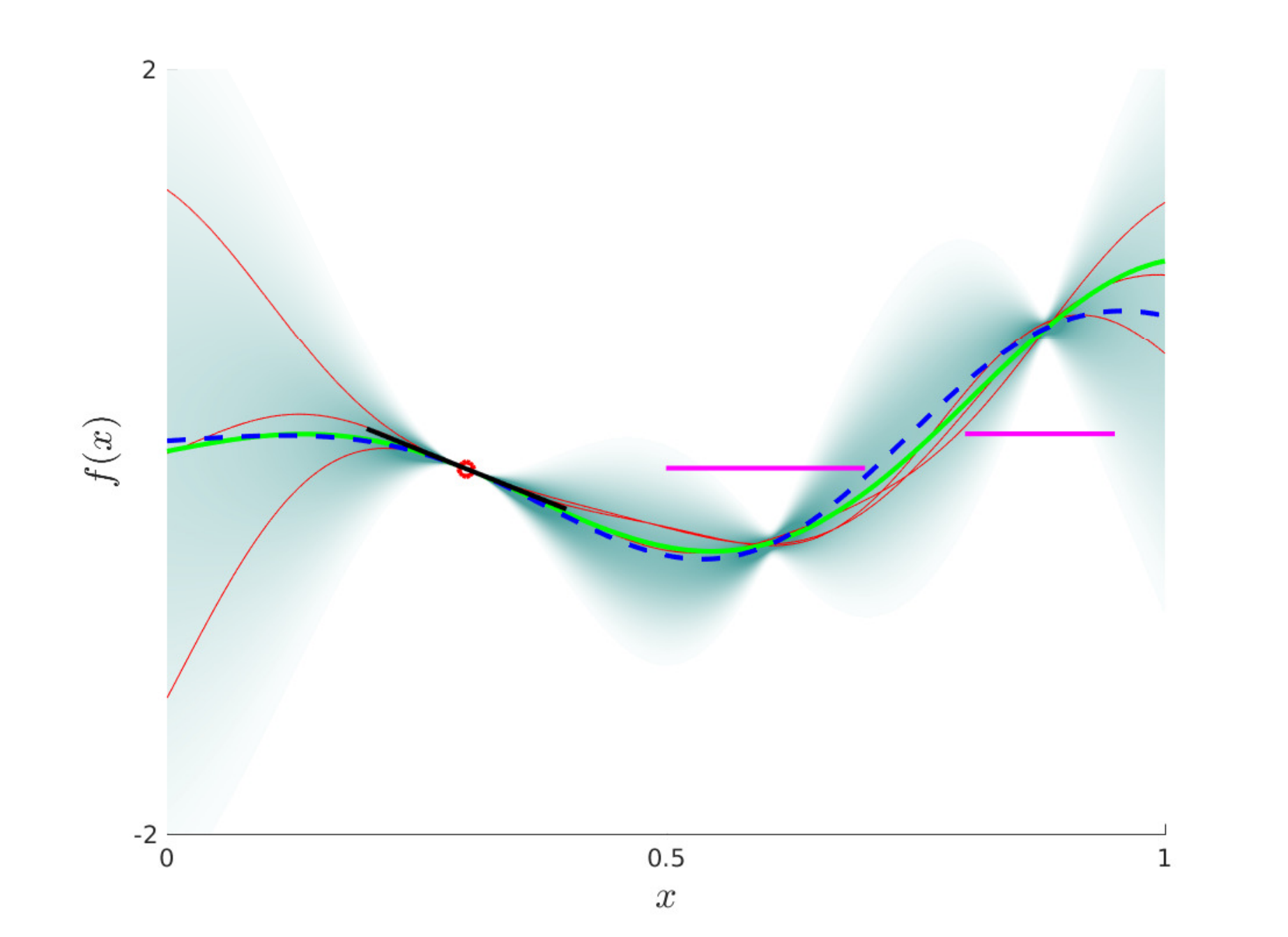}
				\caption{Posterior after one function observation, one derivative observation and two integral observations.}
				\label{fig:funcobs_derint}
			\end{subfigure}
		}
		\caption{Example of GP regression with functional observations. The regression is done using one function observation (red circle), one derivative observation (tangential solid thick black line) and two integral observations (horizontal thick pink lines).}
		\label{fig:funcobs}
	\end{figure}

	An example of GP regression using functional observations is shown in Fig.~\ref{fig:funcobs}, where noise-free observations have been generated from the function $f(x)=x\cos2\pi x$, shown as the solid thick green line.
	The observations consists of one function measurement (red circle), one derivative (tangential solid thick black line) and two integrals (horizontal thick pink lines).
		The squared exponential covariance function $k(x,x')=\sigma_f^2e^{-l^{-2}(x-x')^2/2}$ is used with $\sigma_f=1$ and $l=0.2$.
	Notice that the mean prediction as well as the samples obey the observed properties (although for the integrals this is not directly seen in the plot).
	
	The procedure is easily extended to vector valued functions \citep{Särkkä2011}. 
	Letting $\mathcal{L}_\x$ denote the transformation (such that an element in $\mathcal{L}_\x$ is a linear operator), then we have that
	\begin{equation}\label{eq:vv_transform}
	\mathcal{L}_\x\f
	\sim
	\GP(\mathcal{L}_\x\m(\x),\mathcal{L}_\x\mat{K}_\f(\x,\x')\mathcal{L}_{\x'}^\Transp).
	\end{equation} 
	Since the notation might suggest otherwise, we should emphasise that all operators in $\mathcal{L}_\x\mat{K}(\x,\x')\mathcal{L}_{\x'}^\Transp$ are applied to the elements in $\mat{K}$ from the right. 
	As a simple example, assume that the function $\f(\x):
	\mathbb{R}^2 \mapsto \mathbb{R}^2$ is modelled with a GP
	\begin{align}
	\f(\x) &\sim \GP( 
	\vec{0},\mat{K}_\f(\x,\x')
	),
	\intertext{where}
	\mat{K}_\f(\x,\x')&=
	\begin{bmatrix}
	k_{11}(\x,\x') & k_{12}(\x,\x') \\
	k_{21}(\x,\x') & k_{22}(\x,\x')
	\end{bmatrix}.
	\end{align}
	Assume further that 
	\begin{equation}\label{eq:g_to_f}
	\g =
	\begin{bmatrix}
	\pd{}{x} & 0 \\
	0 & \pd{}{y}
	\end{bmatrix}\f.
	\end{equation}
	Now it follows from \eqref{eq:vv_transform} that $\g$ is also a GP
	\begin{align}
	\nonumber
	\g(\x) &\sim \GP( 
	\m_\g,\mat{K}_\g(\x,\x')
	),
	\intertext{where}
	\m_\g&=				
	\begin{bmatrix}
	\pd{}{x} & 0 \\
	0 & \pd{}{y}
	\end{bmatrix}\vec{0}=\vec{0},
	\end{align}
	and
	\begin{align}
	\mat{K}_\g(\x,\x')&=
	\begin{bmatrix}
	\pd{}{x} & 0 \\
	0 & \pd{}{y}
	\end{bmatrix}
	\begin{bmatrix}
	k_{11}(\x,\x') & k_{12}(\x,\x') \\
	k_{21}(\x,\x') & k_{22}(\x,\x')
	\end{bmatrix}								
	\begin{bmatrix}
	\pd{}{x'} & 0 \\
	0 & \pd{}{y'}
	\end{bmatrix}^\Transp 
	=
	\begin{bmatrix}
	\pdd{}{x}k_{11}(\x,\x') & \pdt{}{x}{y'}k_{12}(\x,\x')\\[1mm]
	\pdt{}{y}{x'}k_{21}(\x,\x') & \pdd{}{y}k_{22}(\x,\x')
	\end{bmatrix}.
	\end{align}
	Using a covariance function with this particular structure ensures that the relation \eqref{eq:g_to_f} is fulfilled.

	\section{Building the covariance function using the Airy stress function}\label{sec:AiryStress}
	To ensure a physical solution, our model should obey the essential \textit{equilibrium constraints}, which must be fulfilled by the strain field at all points.
	As described by \citet[p.~132]{Sadd}, the equilibrium constraints for a two dimensional \textit{stress} field are given by
	\begin{subequations}\label{eq:equi_constr_stress}				\begin{align}
		\frac{\partial \tilde{\sigma}_{xx}}{\partial x}
		+\frac{\partial\sigma_{xy}}{\partial y}=0, \\
		\frac{\partial\sigma_{xy}}{\partial x}
		+\frac{\partial \tilde{\sigma}_{yy}}{\partial y}=0,
		\end{align}
	\end{subequations}
	where $\tilde{\sigma}_{xx}=\sigma_{xx}-V$, $\tilde{\sigma}_{yy}=\sigma_{yy}-V$ and $V$ is a potential function. These equations are satisfied by letting the components be represented as
	\begin{subequations}\label{eq:Sadd752}
		\begin{align}
		\tilde{\sigma}_{xx}&=\frac{\partial^2\varphi}{\partial y^2} ,
		\\
		\sigma_{xy}&=-\frac{\partial^2\varphi}{\partial x\partial y} ,
		\\
		\tilde{\sigma}_{yy}&=\frac{\partial^2\varphi}{\partial x^2} ,
		\end{align}
	\end{subequations} 
	where the arbitrary scalar function $\varphi=\varphi(x,y)$ is the so-called \textit{Airy stress function}. 
	Letting $V=0$ to keep the notation uncluttered, we have that
	\begin{equation}
	\begin{bmatrix}
	\sigma_{xx} \\ 
	\sigma_{xy} \\ 
	\sigma_{yy} 
	\end{bmatrix}
	=
	\begin{bmatrix}
	\frac{\partial^2}{\partial y^2} \\[1mm]
	-\frac{\partial^2}{\partial x\partial y} \\[1mm]
	\frac{\partial^2}{\partial x^2}
	\end{bmatrix}\varphi.
	\end{equation}
	Applying Hooke's law for isotropic linearly elastic solid materials			
	\begin{equation}
	\begin{bmatrix}
	\epsilon_{xx} \\
	\epsilon_{xy} \\ 
	\epsilon_{yy}  
	\end{bmatrix} = \frac{1}{E}
	\begin{bmatrix} 
	1 &  0 & -\nu \\
	0 & 1+\nu & 0 \\
	-\nu & 0 & 1  
	\end{bmatrix}
	\begin{bmatrix}\sigma_{xx} \\ \sigma_{xy} \\ \sigma_{yy} \end{bmatrix},
	\end{equation}
	where $E$ and $\nu$ denote Young's modulus and Poisson's ratio, respectively, we end up with
	\begin{align}\label{eq:calL_app}
	\f=
	\begin{bmatrix}
	\epsilon_{xx} \\
	\epsilon_{xy} \\ 
	\epsilon_{yy}  
	\end{bmatrix} = \frac{1}{E}
	\begin{bmatrix} 
	1 &  0 & -\nu \\
	0 & 1+\nu & 0 \\
	-\nu & 0 & 1  
	\end{bmatrix}
	\begin{bmatrix}
	\frac{\partial^2}{\partial y^2} \\[1mm]
	-\frac{\partial^2}{\partial x\partial y} \\[1mm]
	\frac{\partial^2}{\partial x^2}
	\end{bmatrix}\varphi
	= 
	\frac{1}{E}
	\begin{bmatrix}
	\pdd{}{y}-\nu\pdd{}{x} \\[1mm]
	(1+\nu)\pdt{}{x}{y} \\[1mm]
	\pdd{}{x}-\nu\pdd{}{y}
	\end{bmatrix}\varphi
	=\mathcal{L}_\x\varphi.
	\end{align}
	Let us now model $\varphi$ as a GP 
	\begin{equation}
	\varphi\sim
	\GP(0,k_\varphi(\x,\x')). 
	\end{equation}
	Since the strain field $\f$ is mapped from $\varphi$ through the linear functional $\mathcal{L}_\x$, it follows from Section \ref{sec:func_observs} that it is also a GP
	\begin{equation}
	\f\sim
	\GP(\bs{0},
	\mathcal{L}_\x k_\varphi(\x,\x')\mathcal{L}_{\x'}^\Transp).
	\end{equation} 
	Building the covariance function this way will by construction guarantee that any sample drawn from the resulting posterior fulfils the equilibrium constraints given by \eqref{eq:equi_constr_stress}.

	\section{Details on approximative computations}\label{app:approx}
		\subsection{Elements of the $\vec{Q}$-matrix}
		As stated in \eqref{eq:calL1}, the target function $\f(\x)$ is related to the scalar function $\varphi(x,y)$ according to
		\begin{equation}\label{eq_f_from_phi}
		\begin{split}
		\f(\x)=                                         
		\begin{bmatrix}                                        
		\frac{\partial^2}{\partial y^2}-\nu\frac{\partial^2}{\partial x^2}   \\[1mm]
		(1+\nu)\frac{\partial ^2}{\partial x\partial y}                       \\[1mm]
		\frac{\partial ^2}{\partial x^2}-\nu\frac{\partial ^2}{\partial y^2} 
		\end{bmatrix}\varphi(x,y)
		=
		\mathcal{L}_\x\varphi(x,y).
		\end{split}                                           
		\end{equation}
		In the approximative method described in Section \ref{sec:approx}, we are projecting the functional \eqref{eq:functional_li} onto the basis functions
		\begin{align}
			\phi(\x)  &=\frac{1}{\sqrt{\rho_x\rho_y}}\sin\big(\lambda_{x}(x+\rho_x)\big)\sin\big(\lambda_{y}(y+\rho_y)\big) .
		\end{align}
		To keep the notation uncluttered, we are in this section omitting the indexing of the basis functions as well as the measurements, but keep in mind that each calculation described by the equations below must be repeated $m$ times for each measurement. 
		
		Each measurement with inputs $\x^0$, $L$ and $\hat{n}$ requires us to calculate the integral
		\begin{equation}\label{eq:functional_li_arg}
			\frac{1}{L}\int_{0}^{L} \text{arg}(x^0+sn_x,y^0+sn_y) ds,
		\end{equation}
		where
		\begin{align}\label{eq:basis_func_int_arg_derform}
		\text{arg}(x,y) =
		\oldvec{\vec{n}}^\Transp\mathcal{L}_\x\phi(x,y)= 
		\bigg[
		n_x^2\left(\frac{\partial^2}{\partial y^2}-\nu\frac{\partial^2}{\partial x^2}\right)+
		2n_xn_y(1+\nu)\frac{\partial ^2}{\partial x\partial y}
		+n_y^2\left(\frac{\partial ^2}{\partial x^2}-\nu\frac{\partial ^2}{\partial y^2}\right) 
		\bigg]\phi(x,y).
		\end{align}
		The partial derivatives involved have the following explicit forms 	
		\begin{subequations}\label{eq:basis_func_ders}
			\begin{align}
			\frac{\partial^2}{\partial x^2}&\phi(x,y)=-\lambda_x^2\phi(x,y), 
			\\[2mm]
			\frac{\partial^2}{\partial x\partial y}&\phi(x,y) = 
			\lambda_x\lambda_y\underbrace{\frac{1}{\sqrt{\rho_x\rho_y}}\cos\big(\lambda_x(x+\rho_x)\big)\cos\big(\lambda_y(y+\rho_y)\big)}_{\phi_c(x,y)} ,
			\\
			\frac{\partial^2}{\partial y^2}&\phi(x,y)=-\lambda_y^2\phi(x,y).
			\end{align}
		\end{subequations}	
		Substituting the expressions \eqref{eq:basis_func_ders} into \eqref{eq:basis_func_int_arg_derform} yields
		\begin{align}
		\text{arg}(x,y) = 
		\underbrace{\left[
			n_x^2\left(\nu\lambda_x^2-\lambda_y^2\right)
			+n_y^2\left(\nu\lambda_y^2-\lambda_x^2\right) 
			\right]}_{C_S}\phi(x,y)		
		+\underbrace{2(1+\nu)n_xn_y\lambda_x\lambda_y}_{C_O}\phi_C(x,y). 
		\end{align}
		We can now see that \eqref{eq:functional_li_arg} involves calculation of the two  integrals
		\begin{subequations}\label{eq:basis_unatt_ints}
			\begin{align}
			I_1=\int_{0}^{L} \phi(x+sn_x,y+sn_y) ds
			=\frac{1}{2\sqrt{\rho_x\rho_y}}
			\bigg[
			\frac{1}{\Lambda^-}\sin(\Lambda^-s+B^-)
			-\frac{1}{\Lambda^+}\sin(\Lambda^+s+B^+)
			\bigg]_{s=0}^{s=L}, 
			\end{align}
			\begin{align}
			I_2=\int_{0}^{L} \phi_C(x+sn_x,y+sn_y) ds
			=\frac{1}{2\sqrt{\rho_x\rho_y}}
			\bigg[
			\frac{1}{\Lambda^-}\sin(\Lambda^-s+B^-)
			+\frac{1}{\Lambda^+}\sin(\Lambda^+s+B^+)
			\bigg]_{s=0}^{s=L} ,
			\end{align}
		\end{subequations}
		where we have defined
		\begin{subequations}
			\begin{align}
			\Lambda^\pm&=n_x\lambda_x\pm n_y\lambda_y, 
			\\
			B^\pm&=\lambda_x(x^0+\rho_x)\pm \lambda_y(y^0+\rho_y). 
			\end{align}
		\end{subequations}
		Finally, we end up with
		\begin{equation}\label{eq_noatt_val}
		\frac{1}{L}\int_{0}^{L} \text{arg}(x^0+sn_x,y^0+sn_y)ds=\frac{C_SI_1+C_OI_2}{L}.
		\end{equation}
		Hence, the element $\vec{Q}_{ij}$ in \eqref{eq:appr_expr} is obtained by in the above calculations use the $i$:th basis function with corresponding eigenvalues, and the input arguments of the $j$:th measurement.
		
		\subsection{Marginal Likelihood Expressions}\label{app:ML}
			By replacing $\K_\mathrm{I}$ in \eqref{eq:ML} with the approximation	
			$
			\K_\mathrm{I}\approx
			\vec{Q}^\Transp\bs{\Lambda}\vec{Q},
			$
			and letting $\hat{\Q}=\vec{Q}^\Transp\bs{\Lambda}\vec{Q}+\sigma^2I$,
			we get the following expression for the logarithm of the approximate marginal likelihood
			\begin{align}
			\log p(\y|\{\bs{\eta}_i\},\bs{\theta}) 
			\approx \log \tilde{p}(\y|\{\bs{\eta}_i\},\bs{\theta})
			= -\frac{1}{2}\log \det(\hat{\Q})
			-\frac{1}{2}\y^\Transp \hat{\Q}^{-1}\y
			-\frac{N}{2}\log2\pi.
			\end{align}
			For simplicity, we separate the partial derivative with respect to the noise $\sigma$ and the partial derivatives with respect to the other hyperparameters:
			\begin{subequations}
				\begin{align}
				\pd{}{\theta_i}\log \tilde{p}(\y|\{\bs{\eta}_i\},\bs{\theta})
				=
				-\frac{1}{2}\pd{\log\det(\hat{\Q})}{\theta_i}-\frac{1}{2}\pd{\y^\Transp \hat{\Q}^{-1}\y}{\theta_i},
				\end{align}
				\begin{align}
				\pd{}{\sigma}\log \tilde{p}(\y|\{\bs{\eta}_i\},\bs{\theta})
				=
				-\frac{1}{2}\pd{\log\det (\hat{\Q})}{\sigma}-\frac{1}{2}\pd{\y^\Transp \hat{\Q}^{-1}\y}{\sigma}.
				\end{align}
			\end{subequations}
			Introducing
			$\Z=
			\sigma^2\bs{\Lambda}^{-1}+
			\Q
			\Q^\Transp$, 
			the explicit expressions are:
			\begin{subequations}
				\begin{align}
				\log \det(\hat{\Q})&=
				-(N-m)\log\sigma^2-\log\det(\Z)-\sum_j\log\bs{\Lambda}_{jj},
				\\
				\pd{\log\det(\hat{\Q})}{\theta_i}&=
				-\sum_j\bs{\Lambda}_{jj}\pd{\bs{\Lambda}_{jj}}{\theta_i}
				+\sigma^2\tr{\Z^{-1}\bs{\Lambda}^{-2}\pd{\bs{\Lambda}}{\theta_i}},
				\\
				\pd{\log\det(\hat{\Q})}{\sigma}&=
				-2\frac{N-m}{\sigma}
				-2\sigma\tr{\Z^{-1}\bs{\Lambda}^{-1}},
				\\
				\frac{1}{2}\y^\Transp \hat{\Q}^{-1}\y&=
				\frac{1}{\sigma^2}
				\left(
				\y^\Transp \Q^\Transp\Z^{-1}\Q\y
				\right),
				\\
				\pd{\y^\Transp \hat{\Q}^{-1}\y}{\theta_i}&=
				\y^\Transp \Q^\Transp \Z^{-1}
				\left[
				\bs{\Lambda}^{-2}\pd{\bs{\Lambda}}{\theta_i}
				\right]
				\Z^{-1}\Q\y,
				\\
				\pd{\y^\Transp \hat{\Q}^{-1}\y}{\sigma}&=
				-\frac{2}{\sigma}\y^\Transp\Q^\Transp \Z^{-1}\bs{\Lambda}^{-1}\Z^{-1}\Q\y
				+\frac{2}{\sigma^3}\y^\Transp\y
				-\frac{2}{\sigma^3}\y^\Transp \Q^\Transp \Z^{-1} \Q\y.
				\end{align}
			\end{subequations}


\bibliography{refs}
\bibliographystyle{apalike-url}

\end{document}